\documentclass{article} 
\usepackage{iclr2023_conference,times}

\usepackage{amsmath,amsfonts,bm}

\def\eqref#1{equation~\ref{#1}}

\def\1{\bm{1}}

\DeclareMathAlphabet{\mathsfit}{\encodingdefault}{\sfdefault}{m}{sl}
\SetMathAlphabet{\mathsfit}{bold}{\encodingdefault}{\sfdefault}{bx}{n}

\usepackage{hyperref}
\usepackage{url}

\usepackage[utf8]{inputenc} 
\usepackage[T1]{fontenc}    
\usepackage{hyperref}       
\usepackage{url}            
\usepackage{booktabs}       
\usepackage{amsfonts}       
\usepackage{nicefrac}       
\usepackage{microtype}      
\usepackage{xcolor}         
\usepackage{microtype}
\usepackage{graphicx}
\usepackage{subfigure}
\usepackage{balance}
\usepackage{booktabs} 
\usepackage{amsthm}
\usepackage{array}
\usepackage{graphicx}
\usepackage{clrscode}
\usepackage{subfigure}
\usepackage{multirow}
\usepackage{multicol}
\usepackage{float}
\usepackage{color}
\usepackage{xcolor}
\usepackage{amsopn}
\usepackage{mathrsfs}
\usepackage{mathtools}
\usepackage{amsmath}
\usepackage{booktabs}
\usepackage{arydshln}
\usepackage{hyperref}
\usepackage{blkarray}
\usepackage{enumerate}
\usepackage{courier}
\usepackage{mathrsfs}
\usepackage{rotating}
\usepackage{bm}
\usepackage{wrapfig}
\usepackage{subfigure}
\usepackage{array}
\usepackage{ragged2e}
\usepackage{hyperref}
\usepackage{amsmath}
\usepackage{threeparttable}    
\usepackage{amsthm,amsmath}
\usepackage{mathrsfs}
\usepackage{makecell}
\usepackage{setspace}

\usepackage[ruled,linesnumbered]{algorithm2e}

\SetKwComment{comment}{ $triangleright$ \ }{}
\theoremstyle{definition}
\theoremstyle{theorem}
\newtheorem{mytheory}{Theorem}

\theoremstyle{proof}

\theoremstyle{remark}
\newtheorem*{remark}{Remark}

\title{Recommendation with User Active \\Disclosing Willingness}

\author{Lei Wang\\
Gaoling School of Artificial Intelligence Renmin University of China\\
\texttt{wanglei154@ruc.edu.cn} \\
\And 
Xu Chen\\
Beijing Key Laboratory of Big Data Management and Analysis Methods\\
Gaoling School of Artificial Intelligence Renmin University of China\\
\texttt{xu.chen@ruc.edu.cn} \\
\And
Quanyu Dai, Zhenhua Dong \\
Huawei Noah's Ark Lab \\
\texttt{\{daiquanyu,dongzhenhua\}@huawei.com} \\
}
\begin{document}

\maketitle

\begin{abstract}
Recommender system has been deployed in a large amount of real-world applications, profoundly influencing people's daily life and production.
Traditional recommender models mostly collect as comprehensive as possible user behaviors for accurate preference estimation.
However, considering the privacy, preference shaping and other issues, the users may not want to disclose all their behaviors for training the model.
In this paper, we study a novel recommendation paradigm, where the users are allowed to indicate their ``willingness'' on disclosing different behaviors, and the models are optimized by trading-off the recommendation quality as well as the violation of the user ``willingness''.
More specifically, we formulate the recommendation problem as a multiplayer game, where the action is a selection vector representing whether the items are involved into the model training.
For efficiently solving this game, we design a tailored algorithm based on influence function to lower the time cost for recommendation quality exploration, and also extend it with multiple anchor selection vectors.
We conduct extensive experiments to demonstrate the effectiveness of our model on balancing the recommendation quality and user disclosing willingness.
\end{abstract}

\section{Introduction}

For a long time, the recommender models are developed based on the key assumption of collaborative filtering (CF), where similar users in the past may also behave similarly in the future.
Usually, the similarities between different users are inferred from their behaviors based on either explicit heuristics~\cite{ricci2015recommender} or latent embeddings~\cite{koren2009matrix}.
Thus, more comprehensive user behaviors are the basis for computing more accurate user similarities and obtaining better recommendation quality.

\begin{wrapfigure}{l}{0.5\textwidth}
\includegraphics[width=0.3\textheight]{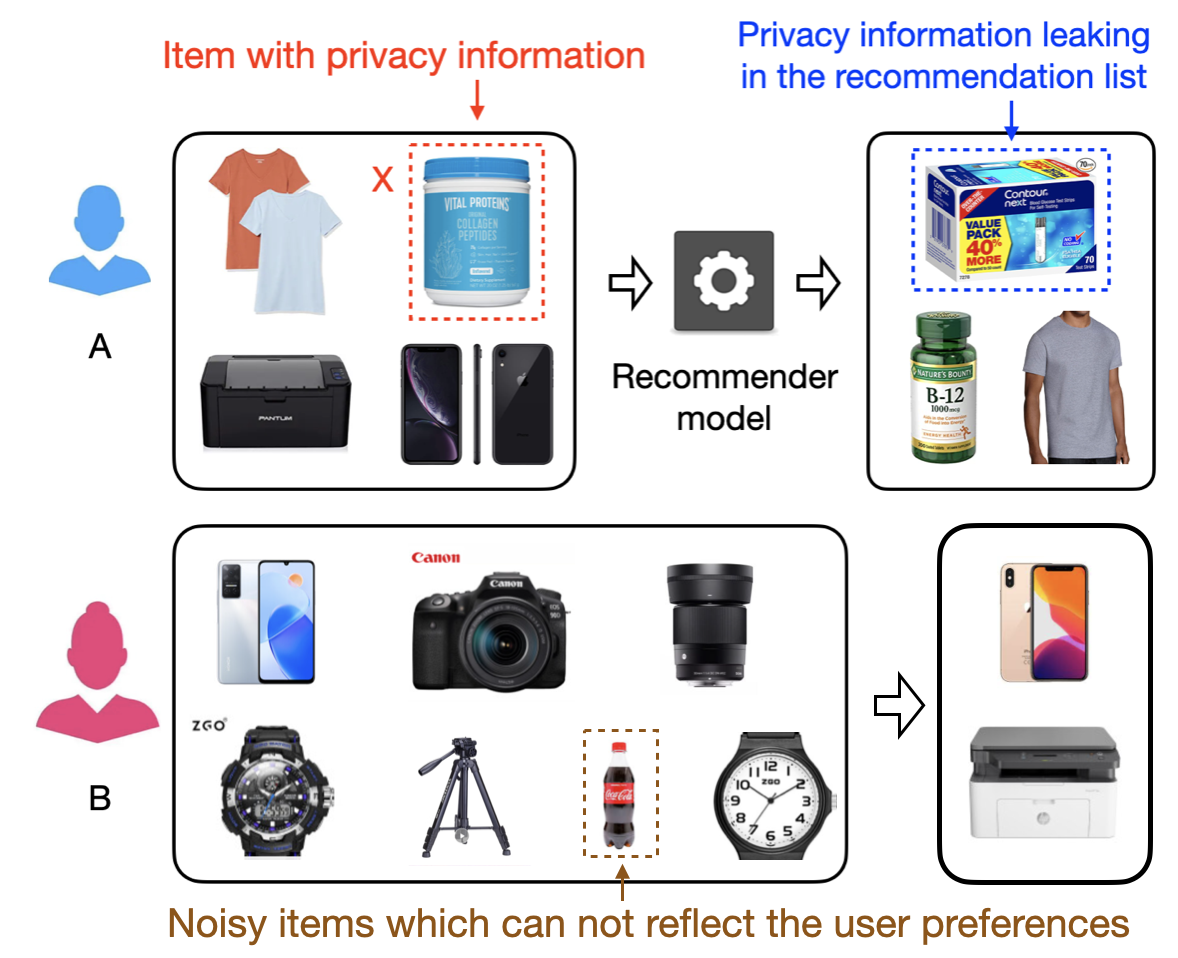}
\vspace{-0.2cm}
\caption{
Motivating examples.
In the first case, the user does not want to disclose her interaction with the medicine.
In the second case, the user would like to remove the items which cannot reflect her preference on the digital items.}
\vspace{-0.4cm}
\label{intro}
\end{wrapfigure}

However, from the perspective of the users, they may not want to disclose all their behaviors for training the model due to the privacy, {preference shaping} and other issues.
As exampled in Figure~\ref{intro}, in the first case, if an item X with privacy information of user A is leveraged to train the model, then the recommendation list may contain items similar to X, and people who have seen this recommendation list may easily infer the user's privacy.
{In the second case, the users may want to actively shape their profiles on the platform by editing their historical interactions, such that the learned model can provide more tailored recommendations.}
Both of the above examples suggest that while more comprehensive user behaviors may promise better user similarity estimation, in real-world scenarios, the users may not want to disclose all of them.

To consider both the recommendation quality and user disclosing willingness, we study a novel recommendation paradigm, where each user can actively specify a willingness vector to show how much she would not like her behaviors to be leveraged for training the recommender model. 
We formulate our problem as a multiplayer game.
Each player corresponds to a user, whose action is a selection vector indicating a subset of the user interacted items\footnote{The interaction with an item can be the user behavior such as click and purchase.}. 
The recommender model is optimized based on all the items selected by different players.
The reward is designed to improve the recommendation quality and simultaneously follow the users' willingness.
To solve this game, the major challenge lies in how to efficiently derive the recommendation quality for different selection vectors.
To overcome this challenge, we firstly set an anchor selection vector, based on which we train the recommender model.
Then, for different actions, we compute the recommendation quality based on the influence function~\cite{koh2017understanding} without retraining the model.
We name our method as \textbf{\underline{i}}nfluence \textbf{\underline{f}}unction based \textbf{\underline{r}}ecommendation \textbf{\underline{q}}uality \textbf{\underline{e}}xploration (called IFRQE for short).
To achieve more accurate quality approximation, we further extend the above method with multiple anchor selection vectors.
We provide theoretical analysis on the designed model in terms of the convergence rate and the benefit of increasing the number of anchor selection vectors.
Extensive experiments based on both synthetic and real-world datasets are conducted to demonstrate the effectiveness of our model.

Notably, there are a few previous works on federated recommendation~\cite{yang2020federated}, recommendation unlearning~\cite{chen2022recommendation} and controllable recommendation.
While these studies share some similarities with our work, they neither allow user active disclosing behaviors nor involve user willingness into the optimization target, which makes them fundamentally differ from our idea.
{~\cite{chen2022proactively} is a recently proposed model for trading-off the user privacy and recommendation quality. Comparing with this work, we improve it by allowing more flexible user willingness, and use influence function to enhance the optimization efficiency.}
In a summary, the main contributions of this paper can be concluded as follows:
(1) We propose a novel recommendation paradigm, where the users can explicitly indicate their disclosing willingness, and the model needs to trade off the recommendation quality and user willingness.
(2) To solve the above problem, we formulate the recommendation task as a multiplayer game, and design two influence function based algorithms to solve the game efficiently.
(3) Extensive experiments are conducted to verify the superiorities of our model based on both synthetic and real-world datasets.

\section{Problem Formulation}
Suppose we have a user set $\mathcal{U}$ and an item set $\mathcal{V}$.
Let $\mathcal{O}^u$ be the set of items interacted by user $u$, and we separate it into a training set $\mathcal{S}^u$, a validation set $\mathcal{T}^u$ and a testing set $\mathcal{D}^u$.
In our problem, each user $u$ can specify a disclosing willingness vector $\bm{\beta}^u = \{\beta^u_1,\beta^u_2,...,\beta^u_{|\mathcal{S}^u|}\}$, where {$\beta^u_k\in [0,1],~\forall k\in [1,|\mathcal{S}^u|]$}.
Suppose $\mathcal{S}^u = \{s_1^u,s_2^u,...,s_{|\mathcal{S}^u|}^u\}$, then the larger $\beta^u_k$ is, the more the user do not want $s_k^u$ to join into the model training.
Our task is to select a subset from $\mathcal{S} = \{\mathcal{S}^u|u\in \mathcal{U}\}$ to train the recommender model, such that the learned model can not only provide acceptable recommendation quality, but also can follow the user willingness.
Intuitively, if we select more items to train the model, then the user can be better understood, and receive higher recommendation quality. 
However, the user willingness is more likely to be violated, since the items with larger ${\beta}_k^u$ may be selected.
Conversely, if we let a small amount of items to train the model, although the user willingness can be better satisfied, the recommendation quality can be lowered.
The key of our task is to learn an optimal strategy, which can well balance the recommendation quality and user willingness.

To solve this task, we regard each user as a player, and formulate the recommendation problem as a multiplayer game.
For each user $u$, the action is a binary selection vector $\bm{o}^u = \{o^u_1,o^u_2,...,o^u_{|\mathcal{S}^u|}\}$, where $o^u_k = 1$ means $s_k^u$ is selected to train the model, otherwise $o^u_k = 0$.
The reward for user $u$ is designed as follows:
\vspace{-0.5cm}
\begin{equation}
\begin{aligned}
\overline{z}_u(\bm{o}^u, \bm{o}^{-u}) = &~-L_f(\mathcal{T}^u,\bm{\hat{\theta}}(\bm{o}^u, \bm{o}^{-u})) - \lambda \sum_{k=1}^{|\mathcal{S}^u|} \bm{o}^u_k {\beta}_k^u\\
\end{aligned}\label{1-reward}
\end{equation}
where $\bm{o}^{-u} = \{\bm{o}^{1},...,\bm{o}^{u-1},\bm{o}^{u+1},...,\bm{o}^{N}\}$ is the joint selection vectors of all the user except $u$.
$f$ is a recommender model.
$\bm{\hat{\theta}}(\bm{o}^u, \bm{o}^{-u})$ are the parameters of $f$ learned based on the training samples selected by $\{\bm{o}^u, \bm{o}^{-u}\}$.
$-L_f(\mathcal{T}^u,\bm{\hat{\theta}}(\bm{o}^u, \bm{o}^{-u}))$ is negative validation loss based on $\bm{\hat{\theta}}(\bm{o}^u, \bm{o}^{-u})$, which is leveraged to measure the recommendation quality.
The second term evaluates the violation of the user willingness. 
If the items that the user does not want to disclose (\emph{e.g.}, ${\beta}_k^u$ is large) are selected to train the model, then $o^u_k{\beta}_k^u$ is large, which lowers the reward.
$\lambda$ is a pre-defined balancing parameter.
Let the strategy of each player be $\bm{\alpha}^u = [\bm{\alpha}^u_{\bm{o}^u}]\in \bigtriangleup(2^{|\mathcal{S}^u|})$, which is a discrete distribution, and $\bm{\alpha}^u_{\bm{o}^u}$ is the probability of leveraging the items indicated by $\bm{o}^u$ to train the model. 
For example, suppose the training set is $\mathcal{S}^u = \{0,1,2\}$, then $\bm{\alpha}^u$ is a distribution defined on $\{\{0,0,0\},\{0,0,1\},\{0,1,0\},\{1,0,0\},\{0,1,1\},\{1,0,1\},\{1,1,0\},\{1,1,1\}\}$, and $\bm{\alpha}^u_{\{0,1,1\}}$ is the probability of using items $\{1,2\}$ to train the model.

We aim to learn the optimal joint strategy $\bm{\alpha}^{*} = \{\bm{\alpha}^{1*}, \bm{\alpha}^{2*},...,\bm{\alpha}^{N*}\}$, such that the corresponding expected reward of each user $u$ is the largest when the other user strategies are fixed, that is:
\begin{equation}
\begin{aligned}
z_u(\bm{\alpha}^{u*}, \bm{\alpha}^{-u*})\geq z_u(\bm{\alpha}^u, \bm{\alpha}^{-u*}),~~\forall u\in [1,2,...,N],~\forall \bm{\alpha}^u \in \bigtriangleup(2^{|\mathcal{S}^u|})
\end{aligned}\label{ne}
\end{equation}
where 
$\bm{\alpha}^{-u} = \{\bm{\alpha}^{1},...,\bm{\alpha}^{u-1},\bm{\alpha}^{u+1},...,\bm{\alpha}^{N}\}$ is the joint strategy of all the users except $u$.
$z_u(\bm{\alpha}^{u}, \bm{\alpha}^{-u})$ is the expected reward for user $u$ under $\{\bm{\alpha}^u, \bm{\alpha}^{-u}\}$, that is, 
$z_u(\bm{\alpha}^{u}, \bm{\alpha}^{-u}) = E_{\bm{o}^u\sim \bm{\alpha}^{u},~\bm{o}^{-u}\sim \bm{\alpha}^{-u}}[\overline{z}_u(\bm{o}^u, \bm{o}^{-u})]$.
After obtaining the optimal strategy $\bm{\alpha}^{*}$, we firstly sample the selection vectors $\bm{o}=\{\bm{o}^u, \bm{o}^{-u}\}$ from $\bm{\alpha}^{*}$.
Then, the training samples are generated by filtering $\mathcal{S}$ with $\bm{o}$, which are leveraged to optimize the final recommender model.
\begin{remark}
(\romannumeral1) The above formulation is from two aspects:
on the one hand, $\beta^u_k$ can be defined based on the privacy concerns, the purpose of reducing the storage/computation burden or any other reasons that the users do not want their behaviors to be disclosed.  
On the other hand, we do not impose constraints on $f$ and $L$, thus they can be any recommender model and loss function, which are compatible with most of the previous work.
(\romannumeral2) In our formulation, we do not assume that the system is adverse. 
The recommender model is developed to better serve the users by following their disclosing willingness and achieving acceptable recommendation quality.
(\romannumeral3) While the action space looks extremely large, which may lower the training efficiency, there are many strategies to alleviate this problem. 
For example, one can assign the same $o^u_k$ for the items in the same category, or only allow the users to indicate their willingness on the most important parts of their interactions.
In addition, we can also initialize $\bm{\alpha}$ with an informative prior, and search the optimal solution around this prior to speed up the training process.
\end{remark}

\section{The IFRQE Model}
\subsection{The basic model}\label{single}
To solve the game defined based on~(\ref{1-reward}) and~(\ref{ne}), one has to derive the validation loss $L_f(\mathcal{T}^u,\bm{\hat{\theta}}(\bm{o}))$ for different $\bm{o}$'s. 
A straightforward method is firstly training $f$ for each $\bm{o}$ to obtain the corresponding parameter $\bm{\hat{\theta}}(\bm{o})$, and then the loss is computed based on the validation set $\mathcal{T}^u$ and $\bm{\hat{\theta}}(\bm{o})$.
However, such method is infeasible, since repeatedly training the recommender model is quite time-consuming.

Fortunately, the previous studies on influence function~\cite{koh2017understanding} may shed some lights on approximating the validation loss without retraining the model.
In specific, we first define an anchor selection vector $\widetilde{\bm{o}} = \{\widetilde{\bm{o}}^{1},...,\widetilde{\bm{o}}^{N}\}$, and then train the recommender model $f$ based on $\widetilde{\bm{o}}$ to obtain the parameters $\widetilde{\bm{\theta}}$.
At last, for a candidate selection vector ${\bm{o}}$, we approximate $L_f(\mathcal{T}^u,\bm{\hat{\theta}}(\bm{o}))$ via the following theory\footnote{We present the proofs of all the theories throughout this paper in the Appendix.}.
\begin{mytheory}
\label{t1}
Given the model parameters $\bm{\theta}$, we define the validation loss by $L_f(\mathcal{T}^u,\bm{{\theta}}) = \sum_{y\in \mathcal{T}^u} l_f(y, {\bm{\theta}})$, where $l_f(y, {\bm{\theta}})$ is the loss of sample $y$ based on ${\bm{\theta}}$.
Suppose
$|\bigtriangledown^2 l_f(s_k^v, \widetilde{\bm{\theta}})|\leq B$ and $\sum_{v}\sum_{k=1}^{|\mathcal{S}^v|} (\widetilde{o}^v_k-o^v_k) B$ is a small value, then the validation loss for $\bm{o}$ is:
\begin{equation}
\begin{aligned}
L_f(\mathcal{T}^u,\bm{\hat{\theta}}(\bm{o})) &\approx L_f(\mathcal{T}^u,\widetilde{\bm{\theta}}) - \frac{1}{Z} \sum_{y\in \mathcal{T}^u}\sum_{v\in \mathcal{U}}\sum_{k=1}^{|\mathcal{S}^v|}o^v_k\bigtriangledown l_f(y,\widetilde{\bm{\theta}})H^{-1}_{\widetilde{\bm{\theta}}} \bigtriangledown l_f(s_k^v, \widetilde{\bm{\theta}}) \\
\end{aligned}\label{3-reward}
\end{equation}
where 
$\widetilde{\bm{\theta}} = \arg\min_{\bm{\theta}} \frac{1}{Z} \sum_{v\in \mathcal{U}}\sum_{k=1}^{|\mathcal{S}^v|}\widetilde{{o}}_{k}^{v} l_f(s_k^v, \bm{\theta})$ are the parameters learned based on the anchor selection vector.
$Z$ is the total number of training samples.
$\bigtriangledown l_f(\cdot)$ is the gradient of a sample loss.
$H_{\widetilde{\bm{\theta}}} = \frac{1}{Z} \sum_{v}\sum_{k=1}^{|\mathcal{S}^v|}\widetilde{{o}}_{k}^{v} \bigtriangledown^2 l_f(s_k^v, \widetilde{\bm{\theta}})$ is the Hessian matrix of the training loss.
\end{mytheory}
Based on this theory, we can compute $L_f(\mathcal{T}^u,\bm{\hat{\theta}}(\bm{o}))$ without retraining the model via $\bm{\hat{\theta}}(\bm{o}) = \arg\min_{{\bm{\theta}}} \frac{1}{Z}\sum_{v}\sum_{k=1}^{|\mathcal{S}^v|}{{o}}^{t,v}_kl_f(s_k^v, {\bm{\theta}})$.
By bringing~(\ref{3-reward}) into~(\ref{1-reward}), $\overline{z}_u(\bm{o}^u, \bm{o}^{-u})$ can be written as:
\begin{equation}
\begin{aligned}
\overline{z}_u(\bm{o}^u, \bm{o}^{-u}) =-L_f(\mathcal{T}^u,\widetilde{\bm{\theta}}) + \frac{1}{Z} \sum_{y\in \mathcal{T}^u}\sum_{v\in \mathcal{U}}\sum_{k=1}^{|\mathcal{S}^v|}o^v_k\bigtriangledown l_f(y,\widetilde{\bm{\theta}})H^{-1}_{\widetilde{\bm{\theta}}} \bigtriangledown l_f(s_k^v, \widetilde{\bm{\theta}}) - \lambda \sum_{k=1}^{|\mathcal{S}^u|} {o}^u_k {\beta}_k^u.\\
\end{aligned}\label{4-reward}
\end{equation}
Accordingly, the expected reward $z_u(\bm{\alpha}^{u}, \bm{\alpha}^{-u})$ is computed via the follow theory.
\begin{mytheory}
Let $\bm{g}^{v}_y = [g(s_1^{v},y),g(s_2^{v},y),...,g(s_{|\mathcal{S}^{v}|}^{v},y)]$, where $g(s_k^v,y) = \bigtriangledown_{\theta}l_f(y,\widetilde{\bm{\theta}})^{T}H_{\widetilde{\bm{\theta}}}^{-1}\bigtriangledown_{\theta} l_f(s_k^v,\widetilde{\bm{\theta}})$, then: $z_u(\bm{\alpha}^{u}, \bm{\alpha}^{-u}) = E_{\bm{o}}[\overline{z}_u(\bm{o}^u, \bm{o}^{-u})] = \sum_{\bm{o}^{u}} {\alpha}^{u}_{\bm{o}^{u}}[(\sum_{y\in \mathcal{T}^u} \frac{\bm{g}^{u}_y}{Z}  - \lambda \bm{\beta}^u)^T\bm{o}^{u}] + C$,
where $C \!= \!-L(\mathcal{T}^u,\widetilde{\bm{\theta}})+ \sum_{v\neq u} E_{\bm{o}}[(\bm{o}^{v})^T\sum_{y\in \mathcal{T}^u} \frac{\bm{g}^{v}_y}{Z}]$ is irrelevant with $\bm{\alpha}^{u}$.
\end{mytheory}
To solve objective~(\ref{ne}), we need to find an optimal $\bm{\alpha}^{u}$ for the following optimization problem:
\begin{equation}
\begin{aligned}
&\max_{\bm{\alpha}^{u}} \sum_{\bm{o}^{u}} {\alpha}^{u}_{\bm{o}^{u}} A(\bm{o}^{u})~~s.t.~\sum_{\bm{o}^u} {\alpha}^{u}_{\bm{o}^{u}} = 1,~~{\alpha}^{u}_{\bm{o}^{u}}\geq 0
\end{aligned}\label{6-obj}
\end{equation}
where we denote $(\sum_{y\in \mathcal{T}^u} \frac{\bm{g}^{u}_y}{N}  - \lambda \bm{\beta}^u)^T\bm{o}^{u}$ by $A(\bm{o}^{u})$.
We present the complete training process of this model in the appendix.

\subsection{The improved model with multiple anchor selection vectors}\label{multi}
The key of the above method lies in the accurate approximation of $L_f(\mathcal{T}^u,\bm{\hat{\theta}}(\bm{o}))$.
However, if the current selection vector $\bm{o}$ is too much different from the anchor vector $\widetilde{\bm{o}}$, then the assumption ``$\sum_{v}\sum_{k=1}^{|\mathcal{S}^v|} (\widetilde{o}^v_k-o^v_k) B$ is a small value in theory~\ref{t1} may not hold, which can lead to larger approximation errors.
To alleviate this problem, we propose to set multiple anchor vectors $\{\widetilde{\bm{o}}^1,\widetilde{\bm{o}}^2,...\widetilde{\bm{o}}^T\}$, where $\widetilde{\bm{o}}^t = \{\widetilde{\bm{o}}^{t,1},...,\widetilde{\bm{o}}^{t,N}\}$ is the $t$th anchor vector and $\widetilde{{o}}_{k}^{t,v}$ is the $k$th element of $\widetilde{\bm{o}}^{t,v}$.
For approximating $L_f(\mathcal{T}^u,\bm{\hat{\theta}}(\bm{o}))$, we select the anchor vector nearest to $\bm{o}$, where we have the following theory.
\begin{mytheory}
For a candidate selection vector $\bm{o}$,
suppose $t = \arg\min_{i\in[1,T]} \sum_{v=1}^N D(\widetilde{\bm{o}}^{i,v}, \bm{o}^{v})$, where $D$ is the hamming distance counting the number of different bits between two vectors, and $\widetilde{\bm{\theta}}^t = \arg\min_{{\bm{\theta}}} \frac{1}{Z}\sum_{v}\sum_{k=1}^{|\mathcal{S}^v|} \widetilde{{o}}^{t,v}_kl_f(s_k^v, {\bm{\theta}})$, then
\vspace{-0.3cm}
\begin{spacing}{0.5}
\begin{equation}
\begin{aligned}
L_f(\mathcal{T}^u,\bm{\hat{\theta}}(\bm{o})) &\approx L_f(\mathcal{T}^u,\widetilde{\bm{\theta}}^t) - \frac{1}{Z} \sum_{y\in \mathcal{T}^u}\sum_{v\in \mathcal{U}}\sum_{k=1}^{|\mathcal{S}^v|}o^v_k\bigtriangledown l_f(y,\widetilde{\bm{\theta}}^t)H^{-1}_{\widetilde{\bm{\theta}}^t} \bigtriangledown l_f(s_k^v, \widetilde{\bm{\theta}}^t) \\
\end{aligned}\label{7-reward}
\end{equation}
\end{spacing}
where $H_{\widetilde{\bm{\theta}}^t} = \frac{1}{Z} \sum_{v}\sum_{k=1}^{|\mathcal{S}^v|} \widetilde{{o}}_{k}^{t,v} \bigtriangledown^2 l_f(s_k^v, \widetilde{\bm{\theta}}^t)$.
\end{mytheory}
Based on~(\ref{7-reward}), the expected reward can be derived based on the following theory.

\setlength{\textfloatsep}{0.2cm}
\begin{algorithm}[t] 
\caption{Learning algorithm for $\bm{\alpha}^{u}_{m}$} 
\label{alg3} 
Indicate the learning rate $\gamma$.\\
Let $\bm{\alpha}^{-u} = \bm{\alpha}_{m-1}^{-u}$ and $\bm{\alpha}^{u}_{1} = \bm{\alpha}^{u}_{m-1}$.\\
\For{l in [1, L]}{
Sample $\bm{o}^{u}, \bm{o}^{-u}$ according to $\bm{\alpha}_{l}^{u}, \bm{\alpha}^{-u}$.\\
Compute the gradient $\hat{g}_{\bm{o}^{u}}$ based on~(\ref{9-app-g}).\\
Let $[\bm{\hat{g}}]_s = {1}(s=\bm{o}^{u}) \hat{g}_{\bm{o}^{u}}$.\\
Update $\bm{\alpha}^{u}_{l+1} = \Pi_{\bigtriangleup}[\bm{\alpha}^{u}_{l} + \gamma \bm{\hat{g}}]$, where $\Pi_{\bigtriangleup}$ means projecting a vector into a simplex.\\
}
Return $\bm{\alpha}^{u} = \frac{1}{L}\sum_{l=1}^L \bm{\alpha}^{u}_{l}$.
\end{algorithm}

\begin{algorithm}[t] 
\caption{Learning Algorithm with Multiple $\theta_t$} 
\label{alg2} 
Initialize $\{\bm{\alpha}^1,\bm{\alpha}^2,...,\bm{\alpha}^N\}$ and let $\bm{\alpha}^u_0 = \bm{\alpha}^u~(u\in[1, N])$.\\
Indicate the max iteration number M and threshold $\kappa$.\\
Indicate the anchor vectors $\{\widetilde{\bm{o}}^1,\widetilde{\bm{o}}^2,...\widetilde{\bm{o}}^T\}$.\\
Obtain the model parameters $\widetilde{\bm{\theta}}^t$ for each $\widetilde{\bm{o}}^t$.\\
\For{m in [1, M]}{
\For{u in [1, N]}{
Let $\bm{\alpha}_{m-1}^{-u} = \{\bm{\alpha}_{m-1}^{1},...,\bm{\alpha}_{m-1}^{u-1},\bm{\alpha}_{m-1}^{u+1},...,\bm{\alpha}_{m-1}^{N}\}$.\\
Obtain $\bm{\alpha}^{u}_{m}$ by inputting $\bm{\alpha}_{m-1}^{-u}$ and $\bm{\alpha}^{u}_{m-1}$ into Algorithm~\ref{alg3}.
}
\If{$|\bm{\alpha}^u_{m}-\bm{\alpha}^u_{m-1}|<\kappa~,\forall u\in[1, N]$}{Break.\\}
}
Output $\bm{\alpha}^{u*} = \bm{\alpha}_m^u~(u\in[1, N])$.
\end{algorithm}

\begin{mytheory}
Let 
${A}_t =\{\bm{o}|\sum_{v=1}^N D(\widetilde{\bm{o}}^{t,v}, \bm{o}^{v}) \leq \sum_{v=1}^N D(\widetilde{\bm{o}}^{t',v}, \bm{o}^{v}),~~\forall t'\neq t\}$.
$g(s_k^{v},y,t) = \bigtriangledown l_f(y,\widetilde{\bm{\theta}}^t)H^{-1}_{\widetilde{\bm{\theta}}^t} \bigtriangledown l_f(s_k^v, \widetilde{\bm{\theta}}^t)$.
$\bm{g}^{t,v}_y = [g(s_1^{v},y,t),g(s_2^{v},y,t),...g(s_{|\mathcal{S}^v|}^{v},y,t)]$ and $\bm{g}^{t,v} = \sum_{y\in \mathcal{T}^u} \bm{g}^{t,v}_y$.
Suppose we define $\overline{z}_u(\bm{o}^u, \bm{o}^{-u},t)= -L_f(\mathcal{T}^u,\widetilde{\bm{\theta}}^t) + \frac{1}{Z} \sum_{y\in \mathcal{T}^u}\sum_{v\in \mathcal{U}}\sum_{k=1}^{|\mathcal{S}^v|}o^v_k\bigtriangledown l_f(y,\widetilde{\bm{\theta}}^t)H^{-1}_{\widetilde{\bm{\theta}}^t} \bigtriangledown l_f(s_k^v, \widetilde{\bm{\theta}}^t)- \lambda \sum_{k=1}^{|\mathcal{S}^u|} {o}^u_k {\beta}_k^u,$
Then 
\vspace{-0.3cm}
\begin{spacing}{0.5}
\begin{equation}
\begin{aligned}
&z_u(\bm{\alpha}^{u}, \bm{\alpha}^{-u}) = E_{\bm{o}}[\overline{z}_u(\bm{o}^u, \bm{o}^{-u})] = E_{\bm{o}}[\sum_{t=1}^T {1}(\bm{o} \in A_t)\overline{z}_u(\bm{o}^u, \bm{o}^{-u},t)] \\
=& \sum_{t=1}^T \sum_{\bm{o}} {1}(\bm{o} \in A_t) \alpha^u_{\bm{o}^u} {\alpha}^{-u}_{\bm{o}^{-u}} \{-L(\mathcal{T}^u, \widetilde{\bm{\theta}}^t) + \frac{1}{Z}\sum_{v\neq u} (\bm{o}^{v})^T \bm{g}^{t,v} + \frac{1}{Z}(\bm{o}^{u})^T \bm{g}^{t,u} - \lambda  (\bm{o}^u)^T \bm{\beta}^u \}\\
\end{aligned}\label{8-expected-reward}
\end{equation}
\end{spacing}
where 
${1}(c) = 1$ if the condition $c$ is true, otherwise ${1}(c) = 0$.
${\alpha}^{-u}_{\bm{o}^{-u}} = {\alpha}^{1}_{\bm{o}^1}...{\alpha}^{(u-1)}_{\bm{o}^{u-1}}{\alpha}^{(u+1)}_{\bm{o}^{u+1}}...{\alpha}^{N}_{\bm{o}^N}$.
\end{mytheory}
Then we derive $\bm{\alpha}^{u}$ by solving the following optimization problem:
\begin{equation}
\begin{aligned}
&\max_{{\alpha}^{u}} \sum_{t=1}^T \sum_{\bm{o}} {1}(\bm{o} \in A_t) \alpha^u_{\bm{o}^u} {\alpha}^{-u}_{\bm{o}^{-u}} B(\bm{o}^u, \bm{o}^{-u}, t)~~~s.t.~\sum_{\bm{o}^u} {\alpha}^{u}(\bm{o}^u) = 1,~~{\alpha}^{u}(\bm{o}^u)>0
\end{aligned}\label{obj-1}
\end{equation}
where 
$B(\bm{o}^u, \bm{o}^{-u}, t) = -L(\mathcal{T}^u,\widetilde{\bm{\theta}}^t) + \frac{1}{Z}\sum_{v\neq u} (\bm{o}^{v})^T \bm{g}^{t,v} + \frac{1}{Z}(\bm{o}^{u})^T \bm{g}^{t,u} - \lambda  (\bm{o}^u)^T \bm{\beta}^u $.

Since it is hard to efficiently obtain a closed-form solution for $\bm{\alpha}^{u}$, we learn it based on the projected gradient descent method~\cite{calamai1987projected}.
More specifically, the gradient of $z_u(\bm{\alpha}^{u}, \bm{\alpha}^{-u})$ w.r.t ${\alpha}^{u}_{\bm{o}^{u}}$ is:
\begin{equation}
\begin{aligned}
\frac{\partial z_u(\bm{\alpha}^{u}, \bm{\alpha}^{-u})}{\partial {\alpha}^{u}_{\bm{o}^{u}}} \overset{\text{def}}{=} g_{\bm{o}^{u}}
&= E_{\bm{o}^{-u}}[\sum_{t=1}^T B(\bm{o}^u, \bm{o}^{-u}, t)\bm{1}(\bm{o}^{-u} \in S_t(\bm{o}^u))].\\
\end{aligned}\label{9-app-g}
\end{equation}
where $S_t(\bm{o}^u) =\{\bm{o}^{-u}|[\bm{o}^u,\bm{o}^{-u}]\in A_t\}$.
Let $\bm{{g}} = [g_{\bm{o}^{u}}]_{\bm{o}^{u}}$ be the gradient of $z_u(\bm{\alpha}^{u}, \bm{\alpha}^{-u})$ w.r.t $\bm{\alpha}^{u}$, which is a $2^{|\mathcal{S}^u|}$ dimensional vector.
$\hat{g}_{\bm{o}^{u}}$ is the stochastic gradient by sampling $\bm{o}^{-u}$ from $\bm{\alpha}^{-u}$.
{Then the complete training process can be seen in Algorithm~\ref{alg2}.}

\subsection{Theoretical analysis}
In this section, we provide theoretical analysis on the convergence of our algorithm and also present the advantage of multiple anchor selection vectors comparing with the single one in terms of the validation loss approximation.
For the convergence analysis, we have the following theory:
\begin{mytheory}
\label{theory5}
Based on the definition of $\bm{{g}}$, we have $z_u(\bm{\alpha}^{u}, \bm{\alpha}^{-u}) = (\bm{\alpha^u})^T{\bm{{g}}}$.
Suppose $\hat{\bm{{g}}}$ is an unbiased estimation of $\bm{{g}}$, and $||\hat{\bm{{g}}}(\bm{\alpha}^{u})||_2^2 \leq G$, then the solution obtained from Algorithm~\ref{alg3} is larger than the optimal solution minus a bounded value, that is: $E[z_u(\bm{\hat{\alpha}}^{u}, \bm{\alpha}^{-u})] \geq \max_{\bm{\alpha}^{u}}E[z_u(\bm{{\alpha}}^{u}, \bm{\alpha}^{-u})] - (\frac{1}{L\gamma} + \gamma^2 G^2)$, 
where $\hat{\bm{\alpha}}^{u}$ is obtained based on Algorithm~\ref{alg3}.
\end{mytheory}
This theory provides foundations for Algorithm~\ref{alg3}, which ensures that the learned strategies are nearly optimal.
For the advantage of using multiple anchor selection vectors, we have the following theory:
\begin{mytheory}
\label{theory6}
For two sets of anchor selection vectors $\bm{P} = \{\widetilde{\bm{o}}^1_P,\widetilde{\bm{o}}^2_P,...\widetilde{\bm{o}}_P^{T_{P}}\}$ and $\bm{Q} = \{\widetilde{\bm{o}}_Q^1,\widetilde{\bm{o}}_Q^2,...\widetilde{\bm{o}}_Q^{T_{Q}}\}$, suppose ${A}^{P}_t=\{\bm{o}|\sum_{v=1}^N D(\widetilde{\bm{o}}^{t,v}, \bm{o}^{v}) \leq \sum_{v=1}^N D(\widetilde{\bm{o}}^{t',v}, \bm{o}^{v}),\forall \widetilde{\bm{o}}^{t'} \in \bm{P} \}$
and
${A}^{Q}_t =\{\bm{o}|\sum_{v=1}^N D(\widetilde{\bm{o}}^{t,v}, \bm{o}^{v}) \leq \sum_{v=1}^N D(\widetilde{\bm{o}}^{t',v}, \bm{o}^{v}),\forall \widetilde{\bm{o}}^{t'} \in \bm{Q}\}$.
Based on theory~\ref{t1}, we consider the following upper bounds of the approximation error for the validation loss:
\vspace{-0.3cm}
\begin{spacing}{0.5}
\begin{equation}
\begin{aligned}
{err}(\bm{P}) = \sum_{t=1}^{T_P} \sum_{\bm{o}\in A^P_t} [\sum_{v} D(\widetilde{\bm{o}}^{t,v}, \bm{o}^{v})]B~~\text{and}~~{err}(\bm{Q}) = \sum_{t=1}^{T_Q} \sum_{\bm{o}\in A^Q_t} [\sum_{v} D(\widetilde{\bm{o}}^{t,v}, \bm{o}^{v})]B,
\end{aligned}
\end{equation}
\end{spacing}
We have if $\bm{P}\!\subseteq\! \bm{Q}$, then ${err}(\bm{P}) \!\!\geq {err}(\bm{Q})$.
\end{mytheory}
This theory suggests if the multiple anchor vectors in section~\ref{multi} are selected to include the single one in section~\ref{single}, then the approximation error upper bound of the validation loss can be reduced.
However, we should note that introducing more anchor vectors means more times for retraining the recommender model, which lowers the efficiency.
Actually, the improved model provides us with an opportunity to balance the approximation accuracy and model efficiency.

\section{Related Work}
Recommender system is a rapidly developing research field, with a large amount of models developed each year~\cite{ricci2015recommender}.
Early recommender models are mostly based on the matrix factorization~\cite{koren2009matrix}.
And then, with the prospering of the deep learning technique, a large amount of deep recommender models~\cite{zhang2019deep} have been proposed, such as the sequential~\cite{kang2018self,tang2018personalized,sun2019bert4rec} or graph~\cite{wang2019kgat,he2020lightgcn,fan2019graph} based algorithms.
These models mostly leverage all the user behaviors to train the model, no matter whether the users would like to disclose them. 
To protect the user privacy, many federated recommender models~\cite{lin2020fedrec,wu2021fedgnn,yang2020federated,chen2020robust} have been proposed.
These studies aim to move the training of the model from the server to the clients, so that the system cannot access the raw user behaviors, and the user privacy is protected. While our model can also be used to protect the privacy, we do not assume that the server is adverse, but aim to ensure that the privacy information does not leak from the recommendation results.
Another topic related to our work is the recommendation unlearning~\cite{chen2022recommendation,li2022making}, which aims to remove the influences of many user-item interactions from the trained recommender model. However, these models do not capture the user active disclosing behaviors, and the user willingness is also not incorporated into the optimization targets.
Many recent studies propose to design controllable recommendation~\cite{wang2022user,parra2015user}, but they mainly aim to break the filter bubbles instead of following the indicated user willingness.
Game theory based recommender models~\cite{ben2018game,xu2018user,halkidi2011game} have also been investigated before. 
For example, \cite{ben2018game} formulates the recommendation problem via a multiplayer game from the content provider perspective.
~\cite{xu2018user,halkidi2011game} study the incentive mechanisms to encourage user rating behaviors and save the user cost simultaneously.
While these models share some similarities with our work, we build the game from the user side and do not limit our framework to the rating behaviors.
In addition, we leverage the influence function to improve the efficiency of recommendation quality estimation, which has not been explored before.
\section{Experiments}
\subsection{Experiment Setup}
\textbf{Datasets.}
We base our experiments on both synthetic and real-world datasets.
For the synthetic dataset, we generated it based on a well known recommendation simulator called RecSim\footnote{https://github.com/google-research/recsim}.
In specific, we generate {1000} users and {1000} items, where {four} features are simulated to profile each user or item. 
For a user-item pair, suppose the user and item features are $\bm{e}_u\in \mathbb{R}^{4}$ and $\bm{e}_v\in \mathbb{R}^{4}$, respectively, then the interaction is generated according to $\bm{1}(\sigma(\bm{e}_u^T\bm{e}_v) \geq \eta)$, where $\bm{1}(\cdot)$ is the indicator function and $\eta$ is a threshold for setting the hardness of generating the interaction.
For the disclosing willingness of user $u$ on item $v$, we simulate it by $\beta^u_v = \sigma(s_u g(\bm{e}_u, \bm{e}_v) + b_u)$, where 
$s_u$ is sampled from a uniform distribution in the range of $[a_1, a_2]$, controlling the sensitivity of the user disclosing willingness.
Smaller $s_u$ means the user has similar willingness on disclosing different behaviors, while larger $s_u$ means the user have more diverse willingness.
$b_u$ is sampled from a Gaussian distribution $\mathcal{N}(0, a_3)$, indicating the average level of the user disclosing willingness.
$g$ is a {one-layer fully connected neural network with randomly assigned parameters}.
For the real-world experiments, we evaluate different models based three public available datasets from different domains, including Diginetica\footnote{https://darel13712.github.io/rs\_datasets/Datasets/diginetica/}, Steam\footnote{https://steam.internet.byu.edu/} and Amazon Video\footnote{https://jmcauley.ucsd.edu/data/amazon/}.
More detailed statistics of these datasets are presented in the Appendix.

\textbf{Baselines.}
To demonstrate the generality of our idea, we implement the base model $f$ with the following recommender algorithms\footnote{Experiments with more other base models can be found in the appendix.}:
{\textbf{MF}~\cite{koren2009matrix} is the traditional matrix factorization method, which has been used as a baseline in a large amount of previous work}.
{\textbf{NeuMF}~\cite{he2017neural} is a well known neural collaborative filtering model, which aims to capture the non-linear correlations between the users and items}.
{\textbf{LightGCN}~\cite{he2020lightgcn} is a graph-based recommender model, which highlights the importance of the structure information in collaborative filtering}.
\textbf{DIN}~\cite{zhou2018deep} is a sequential recommender model, and \textbf{CDAE}~\cite{wu2016collaborative} is an auto-encoder based recommender model.
For each base model, we compare our framework with the following methods,
for the first one, the user behaviors are randomly disclosed (\emph{i.e.}, the selection vector $\bm{o}$ is randomly sampled). We call this method as \textbf{Random}.
For the second one, we directly set $o_u^k=0$, if $\beta_u^k$ is larger than 0.5. We call this method as \textbf{Threshold}.
We also compare our model with \textbf{Proactive}~\cite{chen2022proactively}, which is a model considering both of the recommendation accuracy and user privacy, and the objective is the same as our frameworks'.
The model proposed in section~\ref{single} is named as \textbf{IFRQE}, and the improved version in section~\ref{multi} is called \textbf{IFRQE++}.

\textbf{Implementation Details.}
For each user, we leverage 20\% and 10\% of all her interactions as the testing and validation sets, and the others are left for model training.
For all the models, we use binary cross entropy as the loss function.
To evaluate the recommendation quality, we use $\bm{F_1}$~\cite{chicco2020advantages} as the metric, and {five} items are recommended to compare with the ground truth.
To evaluate whether the model can follow the user willingness, we compute the overall willingness violation by $wv = \frac{1}{|\mathcal{U}|}\sum_{u\in \mathcal{U}}\sum_{k=1}^{|\mathcal{S}^u|} {o}^u_k {\beta}_k^u$, and a model is more satisfied if its $wv$ is smaller.
To evaluate the capability of balancing the recommendation quality and user willingness, we compare different models based on the \textbf{reward} of all the users according to equation (\ref{1-reward}).
It should be noted that our framework and the baselines are optimized with the same reward, and we experiment with different $\lambda$'s to demonstrate the superiority of our framework.
The anchor selection vectors in IFRQE and IFRQE++ are both randomly selected\footnote{We ensure that the anchor selection vectors of IFRQE++ include the one of IFRQE}.
The model hyper parameters are determined by grid search.
For the real world datasets, we specify the disclosing willingness vector of each user in a random manner.
To reduce the randomness, we repeat each experiment for ten times, and report the average performance as well as the standard error.
More settings can be found in the Appendix.

\subsection{Overall Comparison}\label{overall}
The overall comparison results can be seen in Table~\ref{tab:main}, where we can see:
from the perspective of following user willingness, our model can usually perform better than the base model and heuristic methods, which is not surprising since the base model leverages all the items for training, and the heuristic methods consider the user willingness in too coarse manners.
From the recommendation quality perspective, our models do not sacrifice the performance too much.
On average, the performance is dropped by about 8.02\% from the base model.
Interestingly, in some cases, like on the Diginetica and Amazon Video datasets with MF as the base model, the performances are even improved.
We argue that such observation is because many items with higher ${\beta}_k^u$ are exactly not important for the prediction of the items in the testing set.
Thus removing them can not only lead to higher reward via following more user willingness, but also can enhance the performance by focusing on more informative training samples.
From the reward perspective, our model can achieve the best performances on most datasets and base models, which demonstrate the effectiveness and generality of our idea on balancing the recommendation quality and user willingness.
{On average, IFRQE++ can improve the base model by about 12.4\%, 11.1\%, 14.6\% and 21.3\% on the datasets of simulation, Diginetica, Steam and Amazon Video, respectively.}
The improved performance of our models against the random method manifests that the studied problem is non-trivial, where blindly sampling the selection vectors does not work. 
Comparing with Proactive, our model do not improve the performance too much. However, training our framework is, on average, about 13.5 times faster, since Proactive needs to retrain the recommender model frequently.
Between the proposed two methods, the improved version (\emph{i.e.}, {IFRQE++}) is better in more cases.
We speculate that the multiple anchor vectors may lead to more accurate approximation of the validation loss, which may propagate better signals to learn the optimal strategies, and thus result in better rewards.

\begin{table*}[t]
  \caption{{Overall comparison between different models. We use bold fonts to label the best performance for each dataset, evaluation metric and base model. 
  ``()'' indicates the standard error.
  The results of $F_1$ are percentage values with ``\%'' omitted.
  For the metrics, $\uparrow$ means the larger the better, while $\downarrow$ means the lower the better. 
  The performance improvements of our model against the baselines are significant under paired t-test.
  }
  }
  \center
  \renewcommand\arraystretch{1.4}
  \setlength{\tabcolsep}{5.1pt}
  \begin{threeparttable}  
    \scalebox{.65}{
      \begin{tabular}
        { l|
          p{1.1cm}<{\centering}p{1.1cm}<{\centering}p{1.5cm}<{\centering}|
          p{1.1cm}<{\centering}p{1.1cm}<{\centering}p{1.5cm}<{\centering}|
          p{1.1cm}<{\centering}p{1.1cm}<{\centering}p{1.5cm}<{\centering}|
          p{1.1cm}<{\centering}p{1.1cm}<{\centering}p{1.5cm}<{\centering}
        } \hline\hline
        Dataset&
        \multicolumn{3}{c|}{Simulation}&
        \multicolumn{3}{c|}{Diginetica}&
        \multicolumn{3}{c|}{Steam}&
        \multicolumn{3}{c}{Amazon Video}\\ \hline
        Metric&$F_1$ $\uparrow$&$wv$$\downarrow$&reward$\uparrow$&  $F_1$ $\uparrow$&$wv$$\downarrow$&reward$\uparrow$& $F_1$ $\uparrow$&$wv$$\downarrow$&reward$\uparrow$ &$F_1$ $\uparrow$&$wv$$\downarrow$&reward$\uparrow$\\ \hline
        \makecell[c]{MF}  &\textbf{1.75$_{{(.022)}}$}&2.12$_{{(.007)}}$&-2.25$_{{(.032)}}$
        &4.72$_{{(.044)}}$&2.01$_{{(.053)}}$&-2.65$_{{(.072)}}$ 
        &\textbf{10.4}$_{{(.043)}}$&2.62$_{{(.014)}}$&-2.99$_{{(.093)}}$  
        &1.70$_{{(.012)}}$&2.36$_{{(.021)}}$&-2.43$_{{(.022)}}$\\ \hline 
    w/ Random 
        &{1.73}$_{{(.031)}}$&2.01$_{{(.006)}}$&-2.13$_{{(.014)}}$    
        &4.18$_{{(.014)}}$&1.96$_{{(.013)}}$&-2.04$_{{(.006)}}$ 
        &9.40$_{{(.009)}}$&2.61$_{{(.017)}}$&-2.82$_{{(.015)}}$  
        &1.53$_{{(.011)}}$&2.24$_{{(.022)}}$&-2.30$_{{(.023)}}$\\ 
    w/ Threshold
        &{1.70}$_{{(.036)}}$&2.01$_{{(.032)}}$&-2.20$_{{(.021)}}$    
        &4.70$_{{(.012)}}$&1.97$_{{(.011)}}$&-2.06$_{{(.040)}}$ 
        &9.25$_{{(.048)}}$&2.35$_{{(.023)}}$&-2.57$_{{(.017)}}$  
        &1.50$_{{(.026)}}$&2.12$_{{(.027)}}$&-2.20$_{{(.021)}}$\\
    w/ Proactive 
        &{1.23}$_{{(.025)}}$&\textbf{1.56}$_{{(.013)}}$&-2.25$_{{(.007)}}$    
        &2.21$_{{(.018)}}$&\textbf{1.30}$_{{(.024)}}$&-1.99$_{{(.016)}}$ 
        &9.93$_{{(.019)}}$&\textbf{2.08}$_{{(.022)}}$&-2.43$_{{(.025)}}$ 
        &1.55$_{{(.035)}}$&1.94$_{{(.026)}}$&-2.07$_{{(.005)}}$\\
    w/ IFRQE 
        &{1.50}$_{{(.015)}}$&1.97$_{{(.003)}}$&-2.09$_{{(.017)}}$    
        &\textbf{5.23$_{{(.008)}}$}&2.01$_{{(.014)}}$&-2.18$_{{(.006)}}$ 
        &\textbf{10.4$_{{(.012)}}$}&2.14$_{{(.012)}}$&-2.43$_{{(.015)}}$ 
        &1.57$_{{(.015)}}$&\textbf{1.70}$_{{(.016)}}$&-2.10$_{{(.005)}}$\\
    
    w/ IFRQE++ &{1.63}$_{{(.007)}}$&1.82$_{{(.003)}}$&\textbf{-1.92}$_{{(.012)}}$    
                 &{4.02}$_{{(.011)}}$&1.79$_{{(.019)}}$&\textbf{-1.92}$_{{(.022)}}$ 
                 &{9.83}$_{{(.017)}}$&2.13$_{{(.032)}}$&\textbf{-2.42}$_{{(.014)}}$ 
                 &\textbf{1.72}$_{{(.012)}}$&{1.87}$_{{(.032)}}$&\textbf{-1.98}$_{{(.056)}}$\\\hline \hline 

    \makecell[c]{NeuMF} &0.87$_{{(.017)}}$&2.13$_{{(.013)}}$&-2.32$_{{(.011)}}$    
          &\textbf{4.22}$_{{(.013)}}$&2.01$_{{(.015)}}$&-2.11$_{{(.003)}}$ 
          &\textbf{10.2}$_{{(.007)}}$&2.62$_{{(.012)}}$&-2.76$_{{(.007)}}$ 
          &1.65$_{{(.014)}}$&2.36$_{{(.013)}}$&-2.47$_{{(.010)}}$\\ \hline 

    w/ Random  &0.89$_{{(.009)}}$&2.02$_{{(.005)}}$&-2.19$_{{(.011)}}$    
                   &2.38$_{{(.018)}}$&1.91$_{{(.011)}}$&-2.11$_{{(.016)}}$ 
                   &7.73$_{{(.005)}}$&2.47$_{{(.019)}}$&-2.59$_{{(.011)}}$ 
                   &{1.80}$_{{(.014)}}$&2.31$_{{(.006)}}$&-2.41$_{{(.010)}}$\\ 
    w/ Threshold
                   &1.50$_{{(.032)}}$&2.01$_{{(.048)}}$&-2.18$_{{(.031)}}$    
                   &4.01$_{{(.012)}}$&1.97$_{{(.044)}}$&-2.13$_{{(.006)}}$ 
                   &7.50$_{{(.019)}}$&2.35$_{{(.020)}}$&-2.86$_{{(.010)}}$  
                   &1.55$_{{(.002)}}$&2.12$_{{(.018)}}$&-2.21$_{{(.017)}}$\\ 
    w/ Proactive 
        &\textbf{1.83}$_{{(.005)}}$&2.01$_{{(.023)}}$&-2.18$_{{(.017)}}$    
        &1.30$_{{(.008)}}$&\textbf{1.69}$_{{(.011)}}$&-2.14$_{{(.012)}}$ 
        &9.55$_{{(.025)}}$&2.22$_{{(.032)}}$&-2.42$_{{(.005)}}$ 
        &1.43$_{{(.015)}}$&\textbf{1.96}$_{{(.006)}}$&-2.41$_{{(.015)}}$\\
    w/ IFRQE       &1.40$_{{(.014)}}$&1.91$_{{(.013)}}$&-2.48$_{{(.013)}}$    
                   &2.92$_{{(.137)}}$&2.01$_{{(.017)}}$&-2.17$_{{(.019)}}$ 
                   &8.17$_{{(.033)}}$&\textbf{2.16}$_{{(.019)}}$&-2.45$_{{(.027)}}$
                   &\textbf{1.83}$_{{(.015)}}$&1.99$_{{(.014)}}$&-2.24$_{{(.012)}}$\\

    w/ IFRQE++ &{0.70}$_{{(.015)}}$&\textbf{1.80}$_{{(.003)}}$&\textbf{-2.11}$_{{(.011)}}$    
                    &{3.00}$_{{(.017)}}$&1.91$_{{(.013)}}$&\textbf{-2.08}$_{{(.023)}}$ 
                    &{8.87}$_{{(.021)}}$&{2.21}$_{{(.014)}}$&\textbf{-2.37}$_{{(.027)}}$ 
                    &{1.22}$_{{(.031)}}$&{2.00}$_{{(.014)}}$&\textbf{-2.17}$_{{(.011)}}$\\\hline \hline 

    \makecell[c]{LightGCN} &0.90$_{{(.005)}}$&2.13$_{{(.008)}}$&-2.82$_{{(.012)}}$    
                     &\textbf{7.50}$_{{(.009)}}$&2.01$_{{(.012)}}$&-2.71$_{{(.009)}}$ 
                     &10.1$_{{(.010)}}$&2.62$_{{(.009)}}$&-3.13$_{{(.014)}}$ 
                     &\textbf{2.15}$_{{(.021)}}$&2.36$_{{(.008)}}$&-3.05$_{{(.007)}}$\\ \hline 
    
    w/ Random &0.82$_{{(.013)}}$&2.02$_{{(.013)}}$&-2.71$_{{(.022)}}$    
                      &{7.48}$_{{(.011)}}$&1.95$_{{(.005)}}$&-2.64$_{{(.007)}}$ 
                      &10.0$_{{(.027)}}$&2.49$_{{(.026)}}$&-3.07$_{{(.019)}}$ 
                      &1.75$_{{(.003)}}$&2.24$_{{(.009)}}$&-2.93$_{{(.015)}}$\\ 
    w/ Threshold
                      &{0.96}$_{{(.038)}}$&2.01$_{{(.008)}}$&-2.70$_{{(.003)}}$    
                      &7.13$_{{(.017)}}$&1.97$_{{(.040)}}$&-2.60$_{{(.011)}}$ 
                      &9.40$_{{(.048)}}$&2.35$_{{(.027)}}$&-2.82$_{{(.046)}}$  
                      &1.53$_{{(.033)}}$&2.12$_{{(.022)}}$&-2.30$_{{(.048)}}$\\ 
     w/ Proactive 
        &\textbf{1.01}$_{{(.013)}}$&1.78$_{{(.011)}}$&-2.48$_{{(.017)}}$    
        &5.51$_{{(.038)}}$&\textbf{1.42}$_{{(.007)}}$&-2.11$_{{(.018)}}$ 
        &\textbf{10.3$_{{(.021)}}$}&\textbf{2.12}$_{{(.022)}}$&-2.81$_{{(.015)}}$ 
        &1.28$_{{(.013)}}$&1.95$_{{(.016)}}$&-2.64$_{{(.018)}}$\\
    w/ IFRQE  &0.97$_{{(.008)}}$&{2.12}$_{{(.010)}}$&-2.82$_{{(.014)}}$    
                      &4.88$_{{(.019)}}$&1.47$_{{(.011)}}$&-2.16$_{{(.013)}}$ 
                      &9.02$_{{(.016)}}$&2.49$_{{(.025)}}$&-2.82$_{{(.008)}}$ 
                      &1.90$_{{(.007)}}$&\textbf{1.75}$_{{(.018)}}$&-2.13$_{{(.013)}}$\\
    
    w/ IFRQE++ &{0.75}$_{{(.022)}}$&\textbf{1.75}$_{{(.008)}}$&\textbf{-2.44}$_{{(.016)}}$    
                    &{5.26}$_{{(.012)}}$&1.32$_{{(.014)}}$&\textbf{-2.01}$_{{(.009)}}$ 
                    &{8.90}$_{{(.018)}}$&2.48$_{{(.018)}}$&\textbf{-2.80}$_{{(.005)}}$ 
                    &{1.62}$_{{(.013)}}$&\textbf{1.75}$_{{(.009)}}$&\textbf{-2.03}$_{{(.023)}}$\\\hline\hline
    \makecell[c]{DIN} &\textbf{1.43}$_{{(.047)}}$&2.13$_{{(.001)}}$&-2.56$_{{(.044)}}$    
                    &4.13$_{{(.023)}}$&2.01$_{{(.005)}}$&-2.22$_{{(.031)}}$ 
                    &11.6$_{{(.019)}}$&2.62$_{{(.048)}}$&-2.91$_{{(.005)}}$ 
                    &5.05$_{{(.038)}}$&2.36$_{{(.012)}}$&-2.48$_{{(.026)}}$\\ \hline 
   
   w/ Random &1.23$_{{(.024)}}$&2.02$_{{(.037)}}$&-2.42$_{{(.037)}}$    
                     &\textbf{4.33}$_{{(.016)}}$&1.90$_{{(.042)}}$&-2.10$_{{(.024)}}$ 
                     &10.8$_{{(.030)}}$&2.48$_{{(.042)}}$&-2.78$_{{(.014)}}$ 
                     &4.48$_{{(.008)}}$&2.24$_{{(.023)}}$&-2.39$_{{(.015)}}$\\ 
   w/ Threshold
                     &{1.37}$_{{(.004)}}$&2.01$_{{(.022)}}$&-2.13$_{{(.029)}}$    
                     &4.32$_{{(.010)}}$&1.63$_{{(.032)}}$&-1.70$_{{(.036)}}$ 
                     &10.0$_{{(.036)}}$&2.35$_{{(.045)}}$&-2.86$_{{(.029)}}$  
                     &4.25$_{{(.009)}}$&2.12$_{{(.006)}}$&-2.24$_{{(.022)}}$\\ 
     w/ Proactive 
        &{1.21}$_{{(.015)}}$&1.67$_{{(.012)}}$&-2.06$_{{(.007)}}$    
        &2.73$_{{(.008)}}$&1.63$_{{(.024)}}$&-2.19$_{{(.006)}}$ 
        &12.6$_{{(.012)}}$&2.12$_{{(.016)}}$&-2.43$_{{(.015)}}$ 
        &\textbf{5.33}$_{{(.015)}}$&1.84$_{{(.016)}}$&-1.97$_{{(.005)}}$\\
   w/ IFRQE  &1.36$_{{(.030)}}$&{1.10}$_{{(.007)}}$&-1.36$_{{(.016)}}$    
                     &3.92$_{{(.019)}}$&1.46$_{{(.047)}}$&-1.62$_{{(.012)}}$ 
                     &\textbf{13.2}$_{{(.021)}}$&2.57$_{{(.030)}}$&-2.89$_{{(.041)}}$ 
                     &5.26$_{{(.007)}}$&1.61$_{{(.019)}}$&-1.75$_{{(.043)}}$\\
   
   w/ IFRQE++ &{1.03}$_{{(.016)}}$&\textbf{1.09}$_{{(.020)}}$&\textbf{-1.33}$_{{(.045)}}$    
                   &{3.50}$_{{(.014)}}$&\textbf{1.24}$_{{(.042)}}$&\textbf{-1.38}$_{{(.031)}}$ 
                   &{9.53}$_{{(.030)}}$&\textbf{2.02}$_{{(.016)}}$&\textbf{-2.17}$_{{(.045)}}$ 
                   &{5.18}$_{{(.021)}}$&\textbf{1.57}$_{{(.047)}}$&\textbf{-1.70}$_{{(.038)}}$\\\hline\hline
    \makecell[c]{CDAE} &\textbf{1.23}$_{{(.010)}}$&2.13$_{{(.015)}}$&-2.30$_{{(.031)}}$    
                   &\textbf{1.48}$_{{(.009)}}$&2.01$_{{(.036)}}$&-2.08$_{{(.012)}}$ 
                   &11.5$_{{(.008)}}$&2.62$_{{(.029)}}$&-2.68$_{{(.014)}}$ 
                   &1.11$_{{(.011)}}$&2.36$_{{(.027)}}$&-2.44$_{{(.045)}}$\\ \hline 
  
  w/ Random &1.22$_{{(.013)}}$&1.90$_{{(.013)}}$&-2.71$_{{(.022)}}$    
                    &1.38$_{{(.044)}}$&1.99$_{{(.022)}}$&-2.06$_{{(.031)}}$ 
                    &11.6$_{{(.021)}}$&2.49$_{{(.003)}}$&-2.55$_{{(.041)}}$ 
                    &1.21$_{{(.001)}}$&2.24$_{{(.015)}}$&-2.31$_{{(.031)}}$\\ 
  w/ Threshold
                    &0.60$_{{(.021)}}$&2.01$_{{(.003)}}$&-2.08$_{{(.041)}}$    
                    &1.38$_{{(.030)}}$&1.81$_{{(.026)}}$&-1.88$_{{(.016)}}$ 
                    &9.01$_{{(.045)}}$&2.35$_{{(.044)}}$&-2.54$_{{(.010)}}$  
                    &1.22$_{{(.016)}}$&2.12$_{{(.040)}}$&-2.19$_{{(.035)}}$\\ 
     w/ Proactive 
        &{1.06}$_{{(.015)}}$&1.97$_{{(.017)}}$&-2.06$_{{(.027)}}$    
        &1.18$_{{(.008)}}$&1.64$_{{(.011)}}$&-1.76$_{{(.016)}}$ 
        &\textbf{16.3$_{{(.012)}}$}&1.79$_{{(.022)}}$&-1.85$_{{(.025)}}$ 
        &1.63$_{{(.015)}}$&1.79$_{{(.016)}}$&-1.85$_{{(.005)}}$\\
  w/ IFRQE  &0.92$_{{(.011)}}$&1.65$_{{(.015)}}$&-2.07$_{{(.027)}}$    
                    &\textbf{1.48}$_{{(.011)}}$&1.61$_{{(.032)}}$&-1.68$_{{(.013)}}$ 
                    &11.3$_{{(.008)}}$&2.23$_{{(.025)}}$&-2.30$_{{(.023)}}$ 
                    &1.80$_{{(.027)}}$&{1.64}$_{{(.029)}}$&-1.73$_{{(.046)}}$\\
  
  w/ IFRQE++ &{0.86}$_{{(.037)}}$&\textbf{1.45}$_{{(.005)}}$&\textbf{-1.88}$_{{(.012)}}$    
                  &1.43$_{{(.007)}}$&\textbf{1.45}$_{{(.019)}}$&\textbf{-1.52}$_{{(.042)}}$ 
                  &{11.4}$_{{(.018)}}$&\textbf{1.64}$_{{(.046)}}$&\textbf{-1.70}$_{{(.032)}}$ 
                  &\textbf{1.83}$_{{(.049)}}$&\textbf{1.52}$_{{(.009)}}$&\textbf{-1.61}$_{{(.020)}}$\\\hline\hline
      \end{tabular}
    }         
  \end{threeparttable}    
  \label{tab:main}   
\end{table*}

\begin{figure}[t]
  \centering
  \setlength{\fboxrule}{0.pt}
  \setlength{\fboxsep}{0.pt}
  \fbox{
    \includegraphics[width=.75\linewidth]{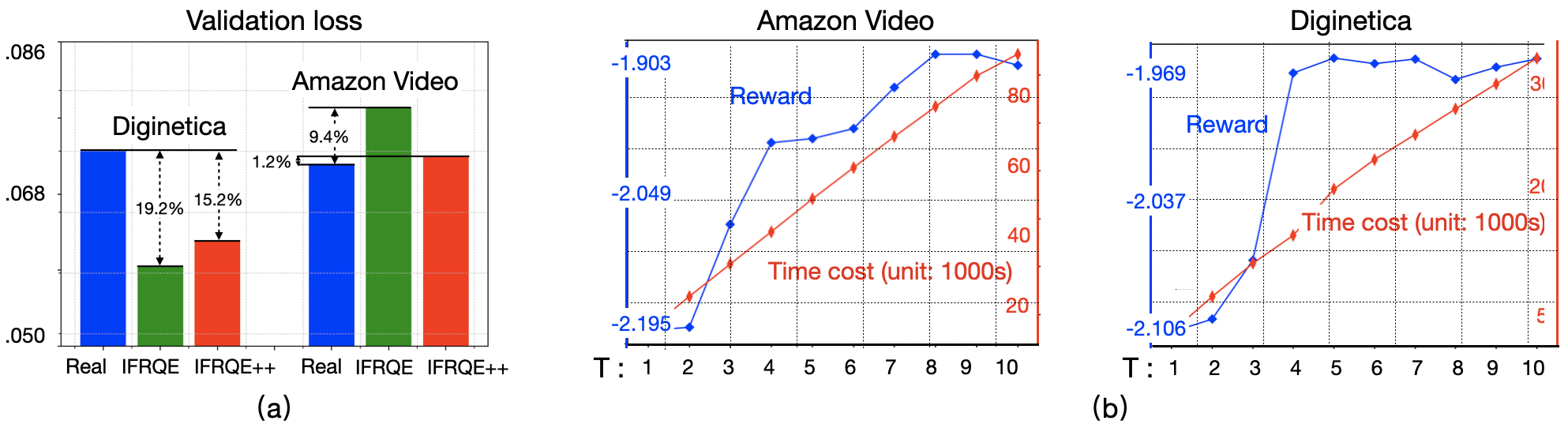}
  }
  \vspace{-0.3cm}
\caption{(a) Approximation error on the validation loss.
(b) Influence of $T$.}
\label{exp}  
\end{figure}

\subsection{Approximation of the Validation Loss}
The key of our model lies in the accurate approximation of the validation loss.
We investigate the approximation error of our models.
We focus on the datasets of {Diginetica and Amazon Video}, and MF is selected as the base model.
We randomly sample ten selection vectors and train the base model to obtain the parameters and real validation loss.
Then, we approximate the validation loss by equation~(\ref{3-reward}) and~(\ref{7-reward}), where we use two anchor vectors in \textbf{IFRQE++}. 
At last, the (approximated) validation losses are reported by averaging over all the selection vectors, and we label the approximation error by $\frac{|\text{real validation loss} - \text{approximated validation loss}|}{\text{real validation loss}}$.
The results are presented in Figure~\ref{exp}(a), and we can see the approximation accuracy of different models is satisfied.
Especially, on Amazon Video, the approximation error of \textbf{IFRQE++} is only 1.2\%.
By leveraging more anchor vectors, \textbf{IFRQE++} is more accurate than \textbf{IFRQE} on both datasets.
In specific, the approximation error is reduced from 19.2\% to 15.2\% and 9.4\% to 1.2\% on the datasets of Diginetica and Amazon Video, respectively.

\subsection{Influence of the Number of Anchor Vectors}
In this section, we study the influence of the number of anchor vectors in \textbf{IFRQE++}.
More specifically, we tune $T$ in the range of $[1,10]$, where $T=1$ is exactly the method proposed in section~\ref{single}.
We base the experiments on the same datasets and base model as in the above experiment.
We report the reward as well as the time cost for each $T$ in Figure~\ref{exp}(b).
We can see: on both datasets, the performances are not satisfied when $T=1$, and more anchor vectors can indeed improve the performance.
With the increasing of $T$, the reward continues to rise up, and tends to be stable at last, where we speculate that, at this time, the performance is dominated by some key anchor vectors, and introducing the other ones do not make much help.
From the efficiency perspective, as $T$ becomes larger, our model costs more time, which is as expected, since larger $T$ means more times to retrain the base model.
In practice, one can properly set $T$ to achieve better trade-offs between the effectiveness and efficiency.

\begin{figure}[t]
  \centering
  \setlength{\fboxrule}{0.pt}
  \setlength{\fboxsep}{0.pt}
  \fbox{
    \includegraphics[width=.75\linewidth]{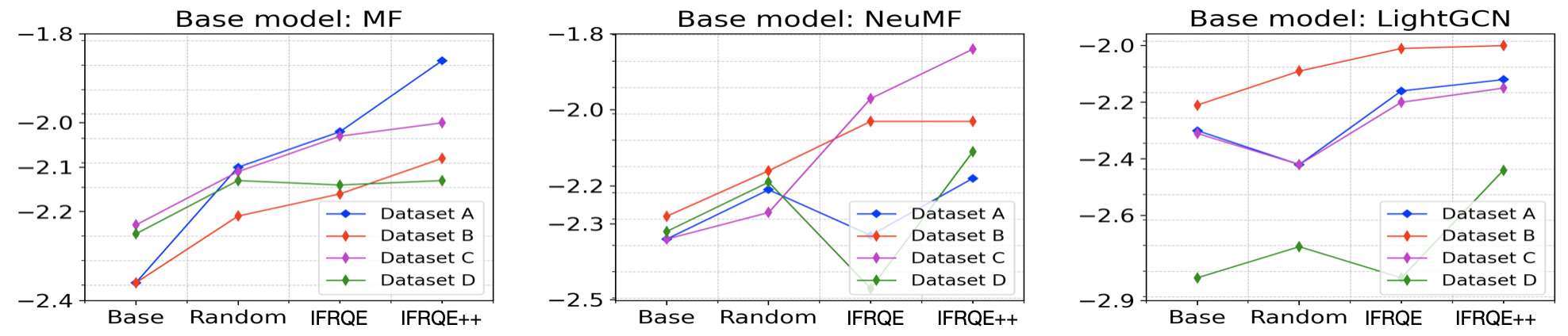}
  }
\vspace{-0.1cm}
\caption{Model comparisons on the datasets with differently characterized user willingness.}
\label{uw}  
\end{figure}
\subsection{Influence of Different Characters of the User Willingness}
Here we study whether our model is always effective on differently characterized user willingness.
We base the experiments on the simulation datasets, where the disclosing willingness of a user is controlled by the sensitive parameter $s_u$ and average disclosing level $b_u$. 
Considering that the distributions of $s_u$ and $b_u$ are determined by $a_1$, $a_2$ and $a_3$, we build four datasets by specifying $(a_1,a_2,a_3)$ with $A = (0.6, 2, 1.2)$, $B = (0.4, 2, 0.8)$ and $C = (0.6, 1, 1.5)$ and $D = (0.5,1,1)$ to simulate different user willingness characters.
We present the comparisons of different models on these datasets in Figure~\ref{uw}, where we can see: 
 the random method is less competitive, and sometimes, it is even worse than the base model (\emph{e.g.}, Datasets A and B with LightGCN as the base model).
For different datasets and base models, IFRQE and IFRQE++ can usually obtain the second best and best performances.
This result agrees with the observations in section~\ref{overall} and manifests that our methods are generally effective for different user willingness characters.

\section{Conclusion and Future Work} 
This paper improves traditional system-centered recommendation paradigm by allowing the users to participate into the model optimization process via specifying their disclosing willingness.
However, there are some limitations can be studied in the future.
We do not consider the dynamic nature of the recommender system, therefore, an interesting direction is to study the temporal user disclosing willingness, for example, under on-line settings.
In addition, we currently assume that the user overall utility is a linear combination between the recommendation quality and user willingness.
One can extend it to the non-linear case, where more efforts are needed to solve the multiplayer game.

\section*{Reproducibility Statement}
For the experimental results presented in the paper, we include the code and simulation dataset in the supplemental
material, and specify all the training details in Appendix. For the datasets used in
the paper, we also give a clear explanation in Section 5.1. More details are available at https://ifrqe.github.io/IFRQE/.
\bibliography{iclr2023_conference}
\bibliographystyle{iclr2023_conference}

\appendix
\section{Appendix}
\subsection{Proof of theory 1}
\begin{proof}
    For a candidate selection vector $\bm{o}$, we have $\hat{\bm{\theta}}(\bm{o}) = \arg\min_{\bm{\theta}} \frac{1}{Z}\sum_{v}\sum_{k=1}^{|\mathcal{S}^v|} o^v_k l_f(s_k^v, \bm{\theta})$, where if $o^v_k=0$, then $s_k^v$ is not leveraged to train the model.
    Suppose we define:
    \begin{equation}
        \begin{aligned}
            \hat{\bm{\theta}}(\bm{o},\epsilon) = \arg\min_{\bm{\theta}} [\frac{1}{Z}\sum_{v}\sum_{k=1}^{|\mathcal{S}^v|} \widetilde{{o}}_{k}^{v}l_f(s_k^v, \bm{\theta}) + \epsilon \sum_{v}\sum_{k=1}^{|\mathcal{S}^v|} (\widetilde{{o}}_{k}^{v}-o^v_k)l_f(s_k^v, \bm{\theta})].
        \end{aligned}\label{supp1}
    \end{equation}
    Then we have $\hat{\bm{\theta}}(\bm{o}) = \hat{\bm{\theta}}(\bm{o},-\frac{1}{Z})$ and $\widetilde{\bm{\theta}} = \hat{\bm{\theta}}(\bm{o}, 0)$.

    Since $\hat{\bm{\theta}}(\bm{o},\epsilon)$ is the optimal solution of~(\ref{supp1}), then
    \begin{equation}
        \begin{aligned}
            0 & \approx \bigtriangledown [\frac{1}{Z} \sum_{v}\sum_{k=1}^{|\mathcal{S}^v|} \widetilde{{o}}_{k}^{v}l_f(s_k^v, \hat{\bm{\theta}}(\bm{o}^u,\epsilon)) + \epsilon \sum_{v}\sum_{k=1}^{|\mathcal{S}^v|} (\widetilde{{o}}_{k}^{v}-o^v_k) l_f(s_k^v, \hat{\bm{\theta}}(\bm{o}^u,\epsilon))]            \\
              & = \frac{1}{Z} \sum_{u}\sum_{k=1}^{|\mathcal{S}^u|} \widetilde{{o}}_{k}^{v}\bigtriangledown l_f(s_k^v, \hat{\bm{\theta}}(\bm{o}^u,\epsilon)) + \epsilon \sum_{v}\sum_{k=1}^{|\mathcal{S}^v|} (\widetilde{{o}}_{k}^{v}-o^v_k) \bigtriangledown l_f(s_k^u, \hat{\bm{\theta}}(\bm{o}^u,\epsilon))]. \\
        \end{aligned}
    \end{equation}
    We regard $\frac{1}{Z} \sum_{v}\sum_{k=1}^{|\mathcal{S}^v|} \widetilde{{o}}_{k}^{v}\bigtriangledown l_f(s_k^v, \bm{\hat{\theta}}(\bm{o},\epsilon)) + \epsilon \sum_{v}\sum_{k=1}^{|\mathcal{S}^v|} (\widetilde{{o}}_{k}^{v}-o^v_k) \bigtriangledown l_f(s_k^u, \bm{\hat{\theta}}(\bm{o},\epsilon))]$ as a function of $\bm{\hat{\theta}}(\bm{o},\epsilon)$.
    If $\epsilon\rightarrow 0$, then $\bm{\hat{\theta}}(\bm{o},\epsilon) \rightarrow \widetilde{\bm{\theta}}$.
    According to the Taylor expansion, we have:
    \begin{equation}
        \begin{aligned}
            0 & \approx \frac{1}{Z} \sum_{u}\sum_{k=1}^{|\mathcal{S}^u|} \widetilde{{o}}_{k}^{v}\bigtriangledown l_f(s_k^u, \widetilde{\bm{\theta}}) + \epsilon \sum_{v}\sum_{k=1}^{|\mathcal{S}^v|} (\widetilde{{o}}_{k}^{v}-o^v_k)\bigtriangledown l_f(s_k^v, \widetilde{\bm{\theta}})                                                             \\
            + & [\frac{1}{Z} \sum_{v}\sum_{k=1}^{|\mathcal{S}^v|} \widetilde{{o}}_{k}^{v}\bigtriangledown^2 l_f(s_k^v, \widetilde{\bm{\theta}}) + \epsilon \sum_{v}\sum_{k=1}^{|\mathcal{S}^v|} (\widetilde{{o}}_{k}^{v}-o^v_k)\bigtriangledown^2 l_f(s_k^v, \widetilde{\bm{\theta}})](\hat{\bm{\theta}}(\bm{o},\epsilon) - \widetilde{\bm{\theta}})
        \end{aligned}
    \end{equation}
    Let $\triangle_{\epsilon} = \hat{\bm{\theta}}(\bm{o},\epsilon) - \widetilde{\bm{\theta}}$, then
    \begin{equation}
        \begin{aligned}
            0                    & \approx \frac{1}{Z} \sum_{v}\sum_{k=1}^{|\mathcal{S}^v|} \widetilde{{o}}_{k}^{v}\bigtriangledown l_f(s_k^v, \widetilde{\bm{\theta}}) + \epsilon \sum_{v}\sum_{k=1}^{|\mathcal{S}^v|} (\widetilde{{o}}_{k}^{v}-o^v_k)\bigtriangledown l_f(s_k^v, \widetilde{\bm{\theta}})                   \\
            +                    & [\frac{1}{Z} \sum_{v}\sum_{k=1}^{|\mathcal{S}^v|} \widetilde{{o}}_{k}^{v}\bigtriangledown^2 l_f(s_k^v, \widetilde{\bm{\theta}}) + \epsilon \sum_{v}\sum_{k=1}^{|\mathcal{S}^v|} (\widetilde{{o}}_{k}^{v}-o^v_k)\bigtriangledown^2 l_f(s_k^v, \widetilde{\bm{\theta}})]\triangle_{\epsilon} \\
            \triangle_{\epsilon} & \approx -[\frac{1}{Z} \sum_{v}\sum_{k=1}^{|\mathcal{S}^v|} \widetilde{{o}}_{k}^{v}\bigtriangledown^2 l_f(s_k^v, \widetilde{\bm{\theta}}) + \epsilon \sum_{v}\sum_{k=1}^{|\mathcal{S}^v|} (\widetilde{{o}}_{k}^{v}-o^v_k)\bigtriangledown^2 l_f(s_k^v, \widetilde{\bm{\theta}})]^{-1}       \\
                                 & [\frac{1}{Z} \sum_{v}\sum_{k=1}^{|\mathcal{S}^v|} \widetilde{{o}}_{k}^{v}\bigtriangledown l_f(s_k^v, \widetilde{\bm{\theta}}) + \epsilon \sum_{v}\sum_{k=1}^{|\mathcal{S}^v|} (\widetilde{{o}}_{k}^{v}-o^v_k)\bigtriangledown l_f(s_k^v, \widetilde{\bm{\theta}})]                         \\
        \end{aligned}\label{cx-app}
    \end{equation}

    Since $|\bigtriangledown^2 l_f(s_k^v, \widetilde{\bm{\theta}})|\leq B$ and $\sum_{v}\sum_{k=1}^{|\mathcal{S}^v|} (\widetilde{o}^v_k-o^v_k) B$ is a small value, we can ignore the term $\epsilon \sum_{v}\sum_{k=1}^{|\mathcal{S}^v|} (\widetilde{{o}}_{k}^{v}-o^v_k)\bigtriangledown^2 l_f(s_k^v, \widetilde{\bm{\theta}})$, which lead to the following equation:
    \begin{equation}
        \begin{aligned}
            \triangle_{\epsilon} & \approx - [\frac{1}{Z} \sum_{v}\sum_{k=1}^{|\mathcal{S}^v|}\widetilde{{o}}_{k}^{v}\bigtriangledown^2 l_f(s_k^v, \widetilde{\bm{\theta}})]^{-1} [\frac{1}{Z} \sum_{v}\sum_{k=1}^{|\mathcal{S}^v|} \widetilde{{o}}_{k}^{v}\bigtriangledown l_f(s_k^v, \widetilde{\bm{\theta}})+\epsilon\sum_{v}\sum_{k=1}^{|\mathcal{S}^v|} (\widetilde{{o}}_{k}^{v}-o^v_k)\bigtriangledown l_f(s_k^v, \widetilde{\bm{\theta}})] \\
        \end{aligned}\label{cx1}
    \end{equation}

    If $Z$ is a large value, then $\frac{d \hat{\bm{\theta}}(\bm{o}, \epsilon)}{d \epsilon}|_{\epsilon \rightarrow 0} \approx \frac{\triangle_{-\frac{1}{Z}}}{-\frac{1}{Z}}$.
    According to~(\ref{cx1}),
    \begin{equation}
        \begin{aligned}
            \triangle_{-\frac{1}{Z}} & \approx - [\frac{1}{Z} \sum_{v}\sum_{k=1}^{|\mathcal{S}^v|}\widetilde{{o}}_{k}^{v} \bigtriangledown^2 l_f(s_k^v, \widetilde{\bm{\theta}})]^{-1} [\frac{1}{Z} \sum_{v}\sum_{k=1}^{|\mathcal{S}^v|} \widetilde{{o}}_{k}^{v}\bigtriangledown l_f(s_k^v, \widetilde{\bm{\theta}})-\frac{1}{Z}\sum_{v}\sum_{k=1}^{|\mathcal{S}^v|} (\widetilde{{o}}_{k}^{v}-o^v_k)\bigtriangledown l_f(s_k^v, \widetilde{\bm{\theta}})] \\
                                     & = - [\frac{1}{Z} \sum_{v}\sum_{k=1}^{|\mathcal{S}^v|}\widetilde{{o}}_{k}^{v}\bigtriangledown^2 l_f(s_k^v, \widetilde{\bm{\theta}})]^{-1} [\frac{1}{Z}\sum_{v}\sum_{k=1}^{|\mathcal{S}^v|} o^v_k\bigtriangledown l_f(s_k^v, \widetilde{\bm{\theta}})]                                                                                                                                                               \\
        \end{aligned}\label{cx2}
    \end{equation}
    Thus, $\frac{d \hat{\bm{\theta}}(\bm{o},\epsilon)}{d \epsilon}|_{\epsilon \rightarrow 0} \approx \frac{\triangle_{-\frac{1}{Z}}}{-\frac{1}{Z}} = [\frac{1}{Z} \sum_{v}\sum_{k=1}^{|\mathcal{S}^v|}\widetilde{{o}}_{k}^{v}\bigtriangledown^2 l_f(s_k^v, \widetilde{\bm{\theta}})]^{-1} [\sum_{v}\sum_{k=1}^{|\mathcal{S}^v|} o^v_k\bigtriangledown l_f(s_k^v, \widetilde{\bm{\theta}})]$

    Because
    \begin{equation}
        \begin{aligned}
            \frac{L_f(\mathcal{T}^u,\bm{\hat{\theta}}(\bm{o})) - L_f(\mathcal{T}^u,\widetilde{\bm{\theta}}) }{-\frac{1}{Z}} & = \frac{L_f(\mathcal{T}^u,\hat{\bm{\theta}}(\bm{o}, -\frac{1}{Z})) - L_f(\mathcal{T}^u,\widetilde{\bm{\theta}}) }{-\frac{1}{Z}} \approx \frac{d L_f(\mathcal{T}^u, \hat{\bm{\theta}}(\bm{o},\epsilon))}{d \epsilon}|_{\epsilon \rightarrow 0} \\
                                                                                                                            & = \sum_{y\in \mathcal{T}^u}\bigtriangledown l_f(y, \widetilde{\bm{\theta}})\times \frac{d \hat{\bm{\theta}}(\bm{o},\epsilon)}{d \epsilon}|_{\epsilon \rightarrow 0}                                                                           \\
                                                                                                                            & \approx \sum_{y\in \mathcal{T}^u}\bigtriangledown l_f(y,\widetilde{\bm{\theta}}) H_{\widetilde{\bm{\theta}}}^{-1} [\sum_{v}\sum_{k=1}^{|\mathcal{S}^v|} o^v_k\bigtriangledown l_f(s_k^v, \widetilde{\bm{\theta}})]                            \\
        \end{aligned}
    \end{equation}
    where $H_{\widetilde{\bm{\theta}}} = \frac{1}{Z} \sum_{v}\sum_{k=1}^{|\mathcal{S}^v|}\widetilde{{o}}_{k}^{v}\bigtriangledown^2 l_f(s_k^v, \widetilde{\bm{\theta}})$, then we have:
    \begin{equation}
        \begin{aligned}
            {L_f(\mathcal{T}^u,\bm{\hat{\theta}}(\bm{o})) - L_f(\mathcal{T}^u,\widetilde{\bm{\theta}}) }
                                                         & \approx {-\frac{1}{Z}} \sum_{y\in \mathcal{T}^u}\bigtriangledown l_f(y,\widetilde{\bm{\theta}}) H_{\widetilde{\bm{\theta}}}^{-1} [\sum_{v}\sum_{k=1}^{|\mathcal{S}^v|} o^v_k\bigtriangledown l_f(s_k^v, \widetilde{\bm{\theta}})]                                      \\
            L_f(\mathcal{T}^u,\bm{\hat{\theta}}(\bm{o})) & \approx L_f(\mathcal{T}^u,\widetilde{\bm{\theta}}) - \frac{1}{Z}\sum_{y\in \mathcal{T}^u}\sum_{v}\sum_{k=1}^{|\mathcal{S}^v|}o^v_k\bigtriangledown l_f(y,\widetilde{\bm{\theta}}) H_{\widetilde{\bm{\theta}}}^{-1}\bigtriangledown l_f(s_k^v, \widetilde{\bm{\theta}}) \\
        \end{aligned}
    \end{equation}
\end{proof}

\subsection{Proof of theory 2}
\begin{proof}
    By bringing the result of theory 1 into $\overline{z}_u(\bm{o}^u, \bm{o}^{-u})$, we have
    \begin{equation}
        \begin{aligned}
            \overline{z}_u(\bm{o}^u, \bm{o}^{-u}) =-L_f(\mathcal{T}^u,\widetilde{\bm{\theta}}) + \frac{1}{Z} \sum_{y\in \mathcal{T}^u}\sum_{v\in \mathcal{U}}\sum_{k=1}^{|\mathcal{S}^v|}o^v_k\bigtriangledown l_f(y,\widetilde{\bm{\theta}})H^{-1}_{\widetilde{\bm{\theta}}} \bigtriangledown l_f(s_k^v, \widetilde{\bm{\theta}}) - \lambda \sum_{k=1}^{|\mathcal{S}^u|} {o}^u_k {\beta}_k^u. \\
        \end{aligned}
    \end{equation}
    Recall that $\bm{g}^{v}_y = [g(s_1^{v},y),g(s_2^{v},y),...,g(s_{|\mathcal{S}^{v}|}^{v},y)]$, where $g(s_k^v,y) = \bigtriangledown_{\theta}l_f(y,\widetilde{\bm{\theta}})^{T}H_{\widetilde{\bm{\theta}}}^{-1}\bigtriangledown_{\theta} l_f(s_k^v,\widetilde{\bm{\theta}})$, then
    \begin{equation}
        \begin{aligned}
              & z_u(\bm{\alpha}^{u}, \bm{\alpha}^{-u}) = E_{\bm{o}}[\overline{z}_u(\bm{o}^u, \bm{o}^{-u})]                                                                                                                                                                           \\
            = & -L_f(\mathcal{T}^u,\widetilde{\bm{\theta}}) + E_{\bm{o}}[\frac{1}{Z} \sum_{y\in \mathcal{T}^u} \sum_{v\in \mathcal{U}} \sum_{k=1}^{|\mathcal{S}^{v}|} o^{v}_k g(s_k^{u'},y) - \lambda \sum_{k=1}^{|\mathcal{S}^u|} o^u_k \beta^u_k]                                  \\
            = & -L_f(\mathcal{T}^u,\widetilde{\bm{\theta}}) + E_{\bm{o}}[\frac{1}{Z}\sum_{y\in \mathcal{T}^u} \sum_{v\in \mathcal{U}} (\bm{o}^{v})^T\bm{g}^{v}_y - \lambda  (\bm{o}^u)^T \bm{\beta}^u]                                                                               \\
            = & -L_f(\mathcal{T}^u,\widetilde{\bm{\theta}}) + E_{\bm{o}}[\sum_{v\in \mathcal{U}} (\bm{o}^{v})^T\sum_{y\in \mathcal{T}^u} \frac{\bm{g}^{v}_y}{Z}  - \lambda  (\bm{o}^u)^T \bm{\beta}^u]                                                                               \\
            = & -L_f(\mathcal{T}^u,\widetilde{\bm{\theta}})+ \sum_{v\neq u} E_{\bm{o}}[(\bm{o}^{v})^T\sum_{y\in \mathcal{T}^u} \frac{\bm{g}^{v}_y}{Z}] + E_{\bm{o}}[(\bm{o}^{u})^T\sum_{y\in \mathcal{T}^u} \frac{\bm{g}^{u}_y}{Z}] - \lambda  E_{\bm{o}}[(\bm{o}^u)^T \bm{\beta}^u] \\
            = & -L_f(\mathcal{T}^u,\widetilde{\bm{\theta}})+ \sum_{v\neq u} E_{\bm{o}}[(\bm{o}^{v})^T\sum_{y\in \mathcal{T}^u} \frac{\bm{g}^{v}_y}{Z}] + E_{\bm{o}^u}[(\bm{o}^{u})^T(\sum_{y\in \mathcal{T}^u} \frac{\bm{g}^{u}_y}{Z} - \lambda \bm{\beta}^u)]                       \\
        \end{aligned}\label{expected-reward-new}
    \end{equation}
\end{proof}

\subsection{Proof of theory 3}
\begin{proof}
    The proof of this theory is similar to that of theory 1.
    We define:
    \begin{equation}
        \begin{aligned}
            \hat{\bm{\theta}}(\bm{o},\epsilon) = \arg\min_{{\bm{\theta}}} [\frac{1}{Z}\sum_{v}\sum_{k=1}^{|\mathcal{S}^v|} \widetilde{\bm{o}}^{t,v}_kl_f(s_k^v, {\bm{\theta}}) + \epsilon \sum_{v}\sum_{k=1}^{|\mathcal{S}^v|} (\widetilde{\bm{o}}^{t,v}_k-o^v_k)l_f(s_k^v, {\bm{\theta}})].
        \end{aligned}
    \end{equation}
    Then,
    \begin{equation}
        \begin{aligned}
            0 & \approx \bigtriangledown [\frac{1}{Z} \sum_{v}\sum_{k=1}^{|\mathcal{S}^v|} \widetilde{{o}}^{t,v}_kl_f(s_k^v, \hat{\bm{\theta}}(\bm{o},\epsilon)) + \epsilon \sum_{v}\sum_{k=1}^{|\mathcal{S}^v|} (\widetilde{{o}}^{t,v}_k-o^v_k) l_f(s_k^v, \hat{\bm{\theta}}(\bm{o},\epsilon))]            \\
              & = \frac{1}{Z} \sum_{u}\sum_{k=1}^{|\mathcal{S}^u|} \widetilde{{o}}^{t,v}_k\bigtriangledown l_f(s_k^v, \hat{\bm{\theta}}(\bm{o},\epsilon)) + \epsilon \sum_{v}\sum_{k=1}^{|\mathcal{S}^v|} (\widetilde{{o}}^{t,v}_k-o^v_k) \bigtriangledown l_f(s_k^u, \hat{\bm{\theta}}(\bm{o},\epsilon))]. \\
        \end{aligned}
    \end{equation}
    According to Taylor expansion at point $\widetilde{\bm{\theta}}^t$, we have:
    \begin{equation}
        \begin{aligned}
            0 & \approx \frac{1}{Z} \sum_{u}\sum_{k=1}^{|\mathcal{S}^u|} \widetilde{{o}}^{t,v}_k\bigtriangledown l_f(s_k^u, \widetilde{\bm{\theta}}^t) + \epsilon \sum_{v}\sum_{k=1}^{|\mathcal{S}^v|} (\widetilde{{o}}^{t,v}_k-o^v_k)\bigtriangledown l_f(s_k^v, \widetilde{\bm{\theta}}^t)                                                               \\
            + & [\frac{1}{Z} \sum_{v}\sum_{k=1}^{|\mathcal{S}^v|} \widetilde{{o}}^{t,v}_k\bigtriangledown^2 l_f(s_k^v, \widetilde{\bm{\theta}}^t) + \epsilon \sum_{v}\sum_{k=1}^{|\mathcal{S}^v|} (\widetilde{{o}}^{t,v}_k-o^v_k)\bigtriangledown^2 l_f(s_k^v, \widetilde{\bm{\theta}}^t)](\hat{\bm{\theta}}(\bm{o},\epsilon) - \widetilde{\bm{\theta}}^t)
        \end{aligned}
    \end{equation}
    Let $\triangle_{\epsilon} = \bm{\hat{\theta}}(\bm{o},\epsilon) - \widetilde{\bm{\theta}}^t$, then:
    \begin{equation}
        \begin{aligned}
            0                    & \approx \frac{1}{Z} \sum_{v}\sum_{k=1}^{|\mathcal{S}^v|} \widetilde{{o}}^{t,v}_k\bigtriangledown l_f(s_k^v, \widetilde{\bm{\theta}}^t) + \epsilon \sum_{v}\sum_{k=1}^{|\mathcal{S}^v|} (\widetilde{{o}}^{t,v}_k-o^v_k)\bigtriangledown l_f(s_k^v, \widetilde{\bm{\theta}}^t)                   \\
            +                    & [\frac{1}{Z} \sum_{v}\sum_{k=1}^{|\mathcal{S}^v|} \widetilde{{o}}^{t,v}_k\bigtriangledown^2 l_f(s_k^v, \widetilde{\bm{\theta}}^t) + \epsilon \sum_{v}\sum_{k=1}^{|\mathcal{S}^v|} (\widetilde{{o}}^{t,v}_k-o^v_k)\bigtriangledown^2 l_f(s_k^v, \widetilde{\bm{\theta}}^t)]\triangle_{\epsilon} \\
            \triangle_{\epsilon} & \approx -[\frac{1}{Z} \sum_{v}\sum_{k=1}^{|\mathcal{S}^v|} \widetilde{{o}}^{t,v}_k\bigtriangledown^2 l_f(s_k^v, \widetilde{\bm{\theta}}^t) + \epsilon \sum_{v}\sum_{k=1}^{|\mathcal{S}^v|} (\widetilde{{o}}^{t,v}_k-o^v_k)\bigtriangledown^2 l_f(s_k^v, \widetilde{\bm{\theta}}^t)]^{-1}       \\
                                 & [\frac{1}{Z} \sum_{v}\sum_{k=1}^{|\mathcal{S}^v|} \widetilde{{o}}^{t,v}_k\bigtriangledown l_f(s_k^v, \widetilde{\bm{\theta}}^t) + \epsilon \sum_{v}\sum_{k=1}^{|\mathcal{S}^v|} (\widetilde{{o}}^{t,v}_k-o^v_k)\bigtriangledown l_f(s_k^v, \widetilde{\bm{\theta}}^t)]                         \\
        \end{aligned}\label{app}
    \end{equation}
    By ignoring $\epsilon \sum_{v}\sum_{k=1}^{|\mathcal{S}^v|} (\widetilde{{o}}^{t,v}_k-o^v_k)\bigtriangledown^2 l_f(s_k^v, \widetilde{\bm{\theta}})$, we have:
    \begin{equation}
        \begin{aligned}
            \triangle_{\epsilon} & \approx - [\frac{1}{Z} \sum_{v}\sum_{k=1}^{|\mathcal{S}^v|} \widetilde{{o}}^{t,v}_k\bigtriangledown^2 l_f(s_k^v, \widetilde{\bm{\theta}}^t)]^{-1} [\frac{1}{Z} \sum_{v}\sum_{k=1}^{|\mathcal{S}^v|} \widetilde{{o}}^{t,v}_k \bigtriangledown l_f(s_k^v, \widetilde{\bm{\theta}}^t)+\epsilon\sum_{v}\sum_{k=1}^{|\mathcal{S}^v|} (\widetilde{{o}}^{t,v}_k-o^v_k)\bigtriangledown l_f(s_k^v, \widetilde{\bm{\theta}}^t)] \\
        \end{aligned}
    \end{equation}
    Then
    \begin{equation}
        \begin{aligned}
            \triangle_{-\frac{1}{Z}} & \approx - [\frac{1}{Z} \sum_{v}\sum_{k=1}^{|\mathcal{S}^v|}\widetilde{{o}}^{t,v}_k\bigtriangledown^2 l_f(s_k^v, \widetilde{\bm{\theta}}^t)]^{-1} [\frac{1}{Z} \sum_{v}\sum_{k=1}^{|\mathcal{S}^v|} \widetilde{{o}}^{t,v}_k\bigtriangledown l_f(s_k^v, \widetilde{\bm{\theta}}^t)-\frac{1}{Z}\sum_{v}\sum_{k=1}^{|\mathcal{S}^v|} (\widetilde{{o}}^{t,v}_k-o^v_k)\bigtriangledown l_f(s_k^v, \widetilde{\bm{\theta}}^t)] \\
                                     & = - [\frac{1}{Z} \sum_{v}\sum_{k=1}^{|\mathcal{S}^v|}\widetilde{{o}}^{t,v}_k\bigtriangledown^2 l_f(s_k^v, \widetilde{\bm{\theta}}^t)]^{-1} [\frac{1}{Z}\sum_{v}\sum_{k=1}^{|\mathcal{S}^v|} o^v_k\bigtriangledown l_f(s_k^v, \widetilde{\bm{\theta}}^t)]                                                                                                                                                                \\
        \end{aligned}\label{cx2}
    \end{equation}
    Thus, $\frac{d \hat{\bm{\theta}}(\bm{o},\epsilon)}{d \epsilon}|_{\epsilon \rightarrow 0} \approx \frac{\triangle_{-\frac{1}{Z}}}{-\frac{1}{Z}} = [\frac{1}{Z} \sum_{v}\sum_{k=1}^{|\mathcal{S}^v|}\widetilde{{o}}^{t,v}_k\bigtriangledown^2 l_f(s_k^v, \widetilde{\bm{\theta}}^t)]^{-1} [\sum_{v}\sum_{k=1}^{|\mathcal{S}^v|} o^v_k\bigtriangledown l_f(s_k^v, \widetilde{\bm{\theta}}^t)]$, and we have
    \begin{equation}
        \begin{aligned}
            \frac{L_f(\mathcal{T}^u,\bm{\hat{\theta}}(\bm{o})) - L_f(\mathcal{T}^u,\widetilde{\bm{\theta}}^t) }{-\frac{1}{Z}} & = \frac{L_f(\mathcal{T}^u,\hat{\bm{\theta}}(\bm{o}, -\frac{1}{Z})) - L_f(\mathcal{T}^u,\widetilde{\bm{\theta}}^t) }{-\frac{1}{Z}} \approx \frac{d L_f(\mathcal{T}^u,\hat{\bm{\theta}}(\bm{o}, \epsilon))}{d \epsilon}|_{\epsilon \rightarrow 0} \\
                                                                                                                              & = \sum_{y\in \mathcal{T}^u}\bigtriangledown l_f(y, \widetilde{\bm{\theta}}^t)\times \frac{d \hat{\bm{\theta}}(\bm{o},\epsilon)}{d \epsilon}|_{\epsilon \rightarrow 0}                                                                           \\
                                                                                                                              & \approx \sum_{y\in \mathcal{T}^u}\bigtriangledown l_f(y,\widetilde{\bm{\theta}}^t) H_{\widetilde{\bm{\theta}}^t}^{-1} [\sum_{v}\sum_{k=1}^{|\mathcal{S}^v|} o^v_k\bigtriangledown l_f(s_k^v, \widetilde{\bm{\theta}}^t)]                        \\
        \end{aligned}
    \end{equation}
    where $H_{\widetilde{\bm{\theta}}^t} = \frac{1}{Z} \sum_{v}\sum_{k=1}^{|\mathcal{S}^v|}\widetilde{{o}}^{t,v}_k\bigtriangledown^2 l_f(s_k^v, \widetilde{\bm{\theta}}^t)$.
    At last,
    \begin{equation}
        \begin{aligned}
            {L_f(\mathcal{T}^u,\bm{\hat{\theta}}(\bm{o})) - L_f(\mathcal{T}^u,\widetilde{\bm{\theta}}^t)}
                                                         & \approx {-\frac{1}{Z}} \sum_{y\in \mathcal{T}^u}\bigtriangledown l_f(y,\widetilde{\bm{\theta}}^t) H_{\widetilde{\bm{\theta}}^t}^{-1} [\sum_{v}\sum_{k=1}^{|\mathcal{S}^v|} o^v_k\bigtriangledown l_f(s_k^v, \widetilde{\bm{\theta}}^t)]                                        \\
            L_f(\mathcal{T}^u,\bm{\hat{\theta}}(\bm{o})) & \approx L_f(\mathcal{T}^u,\widetilde{\bm{\theta}}^t) - \frac{1}{Z}\sum_{y\in \mathcal{T}^u}\sum_{v}\sum_{k=1}^{|\mathcal{S}^v|}o^v_k\bigtriangledown l_f(y,\widetilde{\bm{\theta}}^t) H_{\widetilde{\bm{\theta}}^t}^{-1}\bigtriangledown l_f(s_k^v, \widetilde{\bm{\theta}}^t) \\
        \end{aligned}
    \end{equation}
\end{proof}

\subsection{Proof of theory 4}
\begin{proof}
    \begin{equation}
        \begin{aligned}
              & z_u(\bm{\alpha}^{u}, \bm{\alpha}^{-u}) = E_{\bm{o}}[\overline{z}_u(\bm{o}^u, \bm{o}^{-u})] = E_{\bm{o}}[\sum_{t=1}^T \bm{1}(\bm{o} \in A_t)\overline{z}_u(\bm{o}^u, \bm{o}^{-u},t)]                                                                                                                                              \\
            = & \sum_{t=1}^T \sum_{\bm{o}} \bm{1}(\bm{o} \in A_t)\alpha^u(\bm{o}^u) \bm{\alpha}^{-u}(\bm{o}^{-u}) \{-L_f(\mathcal{T}^u,\widetilde{\bm{\theta}}^t) + \frac{1}{Z}\sum_{y\in \mathcal{T}^u} \sum_{v\in \mathcal{U}} \sum_{k=1}^{|\mathcal{S}^{v}|} o^{v}_k g(s_k^{v},y,t) - \lambda \sum_{k=1}^{|\mathcal{S}^u|} o^u_k \beta^u_k \} \\
            = & \sum_{t=1}^T \sum_{\bm{o}} \bm{1}(\bm{o} \in A_t) \alpha^u(\bm{o}^u) \bm{\alpha}^{-u}(\bm{o}^{-u}) \{-L_f(\mathcal{T}^u,\widetilde{\bm{\theta}}^t) + \frac{1}{Z}\sum_{y\in \mathcal{T}^u} \sum_{v\in \mathcal{U}} (\bm{o}^{v})^T\bm{g}^{t,v}_y - \lambda  (\bm{o}^u)^T \bm{\beta}^u \}                                           \\
            = & \sum_{t=1}^T \sum_{\bm{o}} \bm{1}(\bm{o} \in A_t) \alpha^u(\bm{o}^u) \bm{\alpha}^{-u}(\bm{o}^{-u}) \{-L_f(\mathcal{T}^u,\widetilde{\bm{\theta}}^t) + \frac{1}{Z}\sum_{v\in \mathcal{U}} (\bm{o}^{v})^T \bm{g}^{t,v} - \lambda  (\bm{o}^u)^T \bm{\beta}^u \}                                                                      \\
            = & \sum_{t=1}^T \sum_{\bm{o}} \bm{1}(\bm{o} \in A_t) \alpha^u(\bm{o}^u) \bm{\alpha}^{-u}(\bm{o}^{-u}) \{-L_f(\mathcal{T}^u,\widetilde{\bm{\theta}}^t) + \frac{1}{Z}\sum_{v\neq u} (\bm{o}^{v})^T \bm{g}^{t,v} + \frac{1}{Z}(\bm{o}^{u})^T \bm{g}^{t,u} - \lambda  (\bm{o}^u)^T \bm{\beta}^u \}                                      \\
        \end{aligned}\label{expected-reward-new-1}
    \end{equation}
\end{proof}

\subsection{Proof of theory 5}

We rewrite the objective as follows:
\begin{equation}
    \begin{aligned}
         & \max_{\bm{\alpha}^{u}\in \bigtriangleup}\sum_{\bm{o}^u} \alpha^u_{\bm{o}^u} [\sum_{\bm{o}^{-u}}{\alpha}^{-u}_{\bm{o}^{-u}}\sum_{t=1}^T \bm{1}(\bm{o} \in A_t)B(\bm{o}^u, \bm{o}^{-u}, t)] \\
    \end{aligned}\label{obj-1}
\end{equation}
Suppose the optimal solution for~(\ref{obj-1}) is $\bm{\alpha}^{u}$, and the output of the $l$th iteration is $\bm{\alpha}_l^{u}$.
Recall that $\bm{{g}} =\sum_{\bm{o}^{-u}}{\alpha}^{-u}_{\bm{o}^{-u}}\sum_{t=1}^T \bm{1}(\bm{o} \in A_t)B(\bm{o}^u, \bm{o}^{-u}, t)$, then we have:
\begin{equation}
    \begin{aligned}
        E[\bm{\alpha}^{u}\bm{{g}}-\bm{\alpha}_{l}^{u}\bm{{g}}] & = E[(\bm{\alpha}^{u}-\bm{\alpha}_{l}^{u})\hat{\bm{{g}}}]                                                                                                              \\
                                                               & = E[\frac{1}{2\gamma}(||\bm{\alpha}^{u}-\bm{\alpha}_{l}^{u}||^2_2 + \gamma^2 ||\hat{\bm{g}}||_2^2 - ||\bm{\alpha}^{u}-(\bm{\alpha}_{l}^{u}+\gamma\hat{\bm{g}}||^2_2)] \\
                                                               & \leq E[\frac{1}{2\gamma}(||\bm{\alpha}^{u}-\bm{\alpha}_{l}^{u}||^2_2 + \gamma^2 ||\hat{\bm{g}}||_2^2 - ||\bm{\alpha}^{u}-\bm{\alpha}_{l+1}^{u}||^2_2)]                \\
                                                               & \leq E[\frac{1}{2\gamma}(||\bm{\alpha}^{u}-\bm{\alpha}_{l}^{u}||^2_2 - ||\bm{\alpha}^{u}-\bm{\alpha}_{l+1}^{u}||^2_2 + \gamma^2 G^2 )]                                \\
    \end{aligned}\label{obj-2}
\end{equation}
where the third line hold because $\bm{\alpha}^{u}_{l+1} = \Pi_{\bigtriangleup}[\bm{\alpha}^{u}_{l} + \gamma \bm{\hat{g}}({\alpha}^{u})]$.

In the next,
\begin{equation}
    \begin{aligned}
        \sum_{l=1}^L E[\bm{\alpha}^{u}\bm{y}-\bm{\alpha}_{l}^{u}\bm{y}]
         & \leq E[\frac{1}{2\gamma}(||\bm{\alpha}^{u}-\bm{\alpha}_{1}^{u}||^2_2 - ||\bm{\alpha}^{u}-\bm{\alpha}_{L+1}^{u}||^2_2 + L\gamma^2 G^2 )] \\
         & \leq E[\frac{1}{2\gamma}(||\bm{\alpha}^{u}-\bm{\alpha}_{1}^{u}||^2_2 + L\gamma^2 G^2 )]                                                 \\
         & \leq \frac{1}{\gamma} + L\gamma^2 G^2                                                                                                   \\
    \end{aligned}\label{obj-3}
\end{equation}
where the third line hold because $\bm{\alpha}^{u}$ and $\bm{\alpha}_{1}^{u}$ are both simplex.

At last,
\begin{equation}
    \begin{aligned}
        \frac{1}{L}\sum_{l=1}^L E[\bm{\alpha}^{u}{\bm{g}}-\bm{\alpha}_{l}^{u}{\bm{g}}]      & \leq \frac{1}{L\gamma} + \gamma^2 G^2                                                                   \\
        E[\bm{\alpha}^{u}{\bm{g}}] - \frac{1}{L}\sum_{l=1}^L E[\bm{\alpha}_{l}^{u}{\bm{g}}] & \leq \frac{1}{L\gamma} + \gamma^2 G^2                                                                   \\
        \frac{1}{L}\sum_{l=1}^L E[\bm{\alpha}_{l}^{u}{\bm{g}}]                              & \geq E[\bm{\alpha}^{u}{\bm{g}}] - (\frac{1}{L\gamma} + \gamma^2 G^2)                                    \\
        \frac{1}{L}\sum_{l=1}^L E[\bm{\alpha}_{l}^{u}{\bm{g}}]                              & \geq \max_{\bm{\alpha}^{u}}E[\bm{\alpha}^{u}{\bm{g}}] - (\frac{1}{L\gamma} + \gamma^2 G^2)              \\
        E[z_u(\bm{\hat{\alpha}}^{u}, \bm{\alpha}^{-u})] = E[\hat{\bm{\alpha}}^{u}{\bm{g}}]  & \geq \max_{\bm{\alpha}^{u}}E[\bm{\alpha}^{u}{\bm{g}}] - (\frac{1}{L\gamma} + \gamma^2 G^2)              \\
                                                                                            & = \max_{\bm{\alpha}^{u}}E[z_u(\bm{{\alpha}}^{u}, \bm{\alpha}^{-u})]- (\frac{1}{L\gamma} + \gamma^2 G^2) \\
    \end{aligned}\label{obj-4}
\end{equation}

\subsection{Proof of theory 6}
In equation~(\ref{cx-app}), the approximation error comes from ignoring the term $\sum_{v}\sum_{k=1}^{|\mathcal{S}^v|} (\widetilde{o}^v_k-o^v_k)\bigtriangledown^2 l_f(s_k^v, \widetilde{\bm{\theta}})$, where $\widetilde{o}^v = \{\widetilde{o}^v_1,...\widetilde{o}^v_{|\mathcal{S}^v|}\}$ is the anchor selection vector of user $v$.
Let D be the hamming distance counting the number of different bits between two vectors, then we consider the value $\sum_{v}\sum_{k=1}^{|\mathcal{S}^v|} |\widetilde{o}^v_k-o^v_k|B = \sum_{v} D(\widetilde{\bm{o}}^{v}, \bm{o}^{v})B$, which upper bounds the approximation error, and are interrested in whether the multi-anchor proposal can reduce this value.

In specific, for a given selection vector $\bm{o}$, suppose $\widetilde{\bm{o}}^{t_1}$ is the nearest anchor vector to $\bm{o}$ in $\bm{P}$.
Since $\bm{P}\subseteq \bm{Q}$, we have $\widetilde{\bm{o}}^{t_1}\in \bm{Q}$.
Suppose $\widetilde{\bm{o}}^{t_2}$ is the nearest anchor vector to $\bm{o}$ in $\bm{Q}$, then according the definition of $\bm{Q}$, we have $\sum_{v=1}^N D(\widetilde{\bm{o}}^{t_2,v}, \bm{o}^{v}) \leq \sum_{v=1}^N D(\widetilde{\bm{o}}^{t_1,v}, \bm{o}^{v})$.
Since $B>0$, we have $\sum_{v=1}^N D(\widetilde{\bm{o}}^{t_2,v}, \bm{o}^{v}) B\leq \sum_{v=1}^N D(\widetilde{\bm{o}}^{t_1,v}, \bm{o}^{v})B$.
By summing all the candidate selection vectors, and grouping them according to $A^P_t$ and $A^Q_t$, respectively, we have:
\begin{equation}
    \begin{aligned}
        \sum_{t=1}^{T_P} \sum_{\bm{o}\in A^P_t} [\sum_{v} D(\widetilde{\bm{o}}^{t,v}, \bm{o}^{v})]B \geq \sum_{t=1}^{T_Q} \sum_{\bm{o}\in A^Q_t} [\sum_{v} D(\widetilde{\bm{o}}^{t,v}, \bm{o}^{v})]B,
    \end{aligned}
\end{equation}

\begin{table}[t]
    \centering
    \renewcommand\arraystretch{1.2}
    \caption{{Statistics of the datasets}}
    \scalebox{1.}{
        \begin{threeparttable}
            \begin{tabular}{p{2.8cm}<{\centering}|p{1.6cm}<{\centering}|p{1.6cm}<{\centering}|p{1.8cm}<{\centering}|p{1.8cm}<{\centering}}
                \hline\hline
                Dataset      & \# User & \# Item & \# Interaction & Sparsity \\ \hline
                Simulation   & 1000    & 1000    & 6148           & 99.39\%  \\\hline
                Diginetica   & 2852    & 10739   & 17073          & 99.94\%  \\\hline
                Steam        & 11942   & 6955    & 86595          & 99.89\%  \\\hline
                Amazon Video & 2790    & 12435   & 18703          & 99.95\%  \\\hline
                \hline
            \end{tabular}
        \end{threeparttable}
    }
    \label{rec-dataset}
\end{table}

\begin{figure}[t]
    \centering
    \setlength{\fboxrule}{0.pt}
    \setlength{\fboxsep}{0.pt}
    \fbox{
        \includegraphics[width=1.\linewidth]{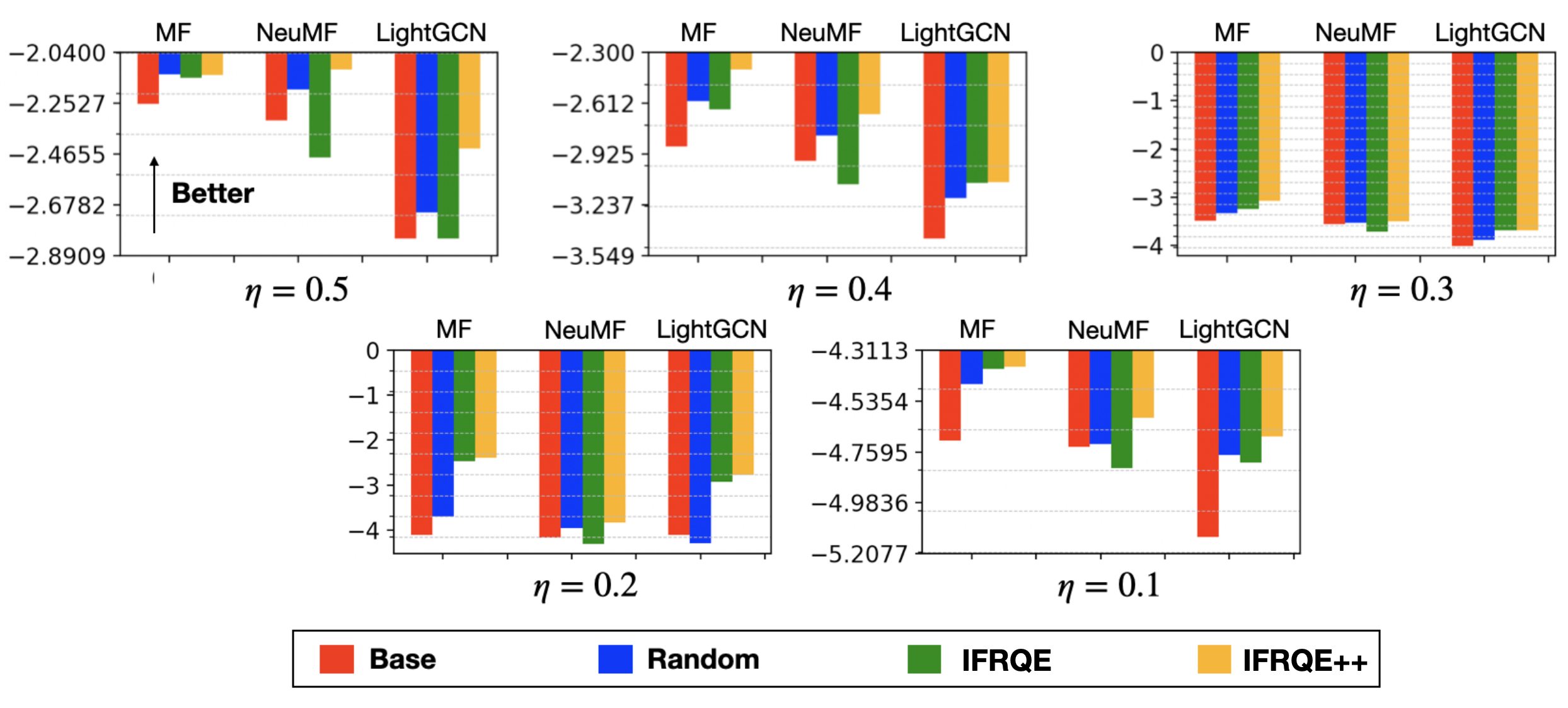}
    }
    \caption{Performance comparison on the dataset with different sparsities.}
    \label{sp}
\end{figure}

\begin{table}[t]
    \centering
    \renewcommand\arraystretch{1.2}
    \caption{{Statistics of the simulation datasets with different $\eta$'s, where the number of users and items are both 1000.}}
    \scalebox{1.}{
        \begin{threeparttable}
            \begin{tabular}{p{2.8cm}<{\centering}|p{1.5cm}<{\centering}|p{1.5cm}<{\centering}|p{1.5cm}<{\centering}|p{1.5cm}<{\centering}|p{1.5cm}<{\centering}}
                \hline\hline
                $\eta$        & 0.1     & 0.2     & 0.3     & 0.4     & 0.5     \\ \hline
                \#Interaction & 11296   & 10068   & 8828    & 7507    & 6184    \\\hline
                Sparsity      & 98.87\% & 99.00\% & 99.12\% & 99.25\% & 99.39\% \\\hline
                \hline
            \end{tabular}
        \end{threeparttable}
    }
    \label{sparse}
\end{table}

\subsection{Learning Algorithm for the Basic Model}
Here, we present the learning algorithms for the basic model in algorithm~\ref{alg1}.

\setlength{\textfloatsep}{0.2cm}
\begin{algorithm}[t]
    \caption{Training process of the based model}
    \label{alg1}
    Initialize $\{\bm{\alpha}^1,\bm{\alpha}^2,...,\bm{\alpha}^N\}$ and let $\bm{\alpha}^u_0 = \bm{\alpha}^u~(u\in[1, N])$.\\
    Indicate the max iteration number M and threshold $\kappa$.\\
    \For{m in [1, M]}{
    \For{u in [1, N]}{
        Let $\bm{\alpha}_{m-1}^{-u} = \{\bm{\alpha}_{m-1}^{1},...,\bm{\alpha}_{m-1}^{u-1},\bm{\alpha}_{m-1}^{u+1},...,\bm{\alpha}_{m-1}^{N}\}$.\\
        Learning $\bm{\alpha}_{m}^{u}$ based on~(\ref{6-obj}) by fixing $\bm{\alpha}_{m-1}^{-u}$.\\
    }
    \If{$|\bm{\alpha}^u_{m}-\bm{\alpha}^u_{m-1}|<\kappa~,\forall u\in[1, N]$}{Break.\\}
    }
    Output $\bm{\alpha}^{u*} = \bm{\alpha}_m^u~(u\in[1, N])$.
\end{algorithm}

\subsection{More Implementation Details}
For the simulation dataset, the threshold $\eta$ and $(a_1, a_2, a_3)$ are initially set as 0.5 and $(0.5, 1, 1)$, respectively.
And then, we tune them in the experiments to study the influence of different dataset sparsities and user willingness characters.
For the real world datasets,
\textbf{Diginetica} and \textbf{Amazon Video} are e-commerce datasets, where we are provided with the user-item purchasing records.
\textbf{Steam} is a game dataset, which includes the interactions (\emph{e.g.}, reviewing behaviors) between the users and games.
Since we do not know the real user disclosing willingness, we simulate it by randomly assigning the willingness vector for each user, and repeat the experiments for ten times to make sure that the experiment results are not from the randomness.
The statistics of the above datasets are concluded in Table~\ref{rec-dataset}.

In IFRQE++, considering that the space of $\bm{\alpha}$ can be extremely large, it is less efficient to initialize $\bm{\alpha}$ completely at random, and blindly learn it in the optimization process.
To solve this problem, we initialize $\bm{\alpha}$ with a prior, assuming that most of the items should be leveraged to train the model for achieving acceptable recommendation performance.
In specific, for each $\bm{\alpha}^u\in \bm{\alpha}$, we initialize it with a Binomial distribution $p(k,n) = C_{n}^k s^k(1-s)^{n-k}$, where $k$ is the number of disclosed items, and $n$ is the total number of items in the training set (\emph{i.e.}, $|\mathcal{S}^u|$).
Notably, we do not discriminate the item differences in the initialization process of $\bm{\alpha}^u$.
For example, suppose there are three items, then $\bm{\alpha}^u_{\{1,1,0\}} = \bm{\alpha}^u_{\{1,0,1\}} = \bm{\alpha}^u_{\{0,1,1\}}$.
In the experiment, we set $s=0.9$, which means, in the beginning, about $0.9*|\mathcal{S}^u|$ items will be involved into the model training process.

In order to efficiently compute the inverse of the Hessian matrix, we use the stochastic estimation method discussed in~\cite{koh2017understanding}.
In specific, according to the Taylor expansion, we can express $H^{-1}$ by $\sum_{i=0}^{\infty}(I-H)^{i}$.
Let $H_j^{-1} = \sum_{i=0}^j(I-H)^{i}$, then we have $H_j^{-1} = I + (I-H)H_{j-1}^{-1}$.
To compute $H^{-1}_{\widetilde{\bm{\theta}}} \bigtriangledown l_f(s_k^v, \widetilde{\bm{\theta}})$, we uniformly sample a training data $s$, and approximate $H$ by $\bigtriangledown^2 l_f(s, \widetilde{\bm{\theta}})$.
Then we have the following recursive equation:
\begin{equation}
    \begin{aligned}
        H^{-1}_j \bigtriangledown l_f(s_k^v, \widetilde{\bm{\theta}}) = \bigtriangledown l_f(s_k^v, \widetilde{\bm{\theta}})  + (I-\bigtriangledown^2 l_f(s, \widetilde{\bm{\theta}}))H^{-1}_{j-1} \bigtriangledown l_f(s_k^v, \widetilde{\bm{\theta}})
    \end{aligned}
\end{equation}
Obviously, when $j\rightarrow \infty$, we have $H^{-1}_j \bigtriangledown l_f(s_k^v, \widetilde{\bm{\theta}}) \rightarrow H^{-1}_{\widetilde{\bm{\theta}}} \bigtriangledown l_f(s_k^v, \widetilde{\bm{\theta}})$.
In the experiment, we resample $s$ for each iteration, and the total number of iterations $N_J$ is tuned to better effectiveness-efficiency trade-off.

For the model parameters, we determine them by grid search.
For example, the number of anchor selection vectors is searched in $[1,2,3,4,5,6,7,8,9,10]$.
The learning rate and batch size are determined in the ranges of $[0.001,0.01,0.05]$ and $[1024,2048,4096]$, respectively.
The anchor selection vectors are sampled from the Binomial distribution, where, similar to $\bm{\alpha}^u$, we set the mean as 0.9.
The final parameters used in our experiments are concluded in Table~\ref{parameter}.
Our project has been released at {https://ifrqe.github.io/IFRQE/}.

\begin{table}[t]
    \centering
    \renewcommand\arraystretch{1.4}
    \caption{Parameter settings in the experiments.}
    \scalebox{.8}{
        \begin{threeparttable}
            \begin{tabular}{p{3.8cm}<{\centering}|p{3.5cm}<{\centering}|p{1.5cm}<{\centering}|p{1.6cm}<{\centering}|p{1.6cm}<{\centering}|p{2.4cm}<{\centering}}
                \hline\hline
                Parameter                    & Tuning range                        & Simulation & Diginetica & Steam & Amazon Video \\ \hline
                Learning rate                & $\left[0.001,0.01,0.05\right]$      & 0.01       & 0.01       & 0.01  & 0.01         \\\hline
                Batch size                   & $[1024,2048,4096]$                  & 2048       & 2048       & 2048  & 2048         \\\hline
                Embedding size               & $\left[64,128,256\right]$           & 64         & 64         & 64    & 64           \\\hline
                Drop ratio                   & $\left[0.01,0.1,0.2\right]$         & 0.1        & 0.1        & 0.1   & 0.1          \\\hline
                $\lambda$                    & $\left[0.1,0.5,1\right]$            & 1          & 1          & 1     & 1            \\\hline
                Iteration number M           & $\left[1,3,5,10\right]$             & 10         & 10         & 10    & 10           \\\hline
                Training epochs              & $\left[50,100,150\right]$           & 50         & 50         & 150   & 100          \\\hline
                L                            & $\left[500,1000,2000\right]$        & 1000       & 1000       & 500   & 1000         \\\hline
                T                            & $\left[1,2,3,4,5,6,7,8,9,10\right]$ & 2          & 8          & 4     & 6            \\\hline
                $N_J$ for computing $H^{-1}$ & $\left[10,20,30\right]$             & 30         & 10         & 20    & 20           \\\hline
                \hline
            \end{tabular}
        \end{threeparttable}
    }
    \label{parameter}
\end{table}

\begin{table*}[h]
    \caption{{Comparison between between our framework and the method of training from scratch (SCR).
                Results based on base models MF, NeuMF, LightGCN, DIN and CDAE.
                ``()'' indicates the standard error.
                The metrics for evaluating the recommendation performance are percentage values with ``\%'' omitted.
                For the metrics, $\uparrow$ means the larger the better, while $\downarrow$ means the lower the better.
            }}
    \center
    \renewcommand\arraystretch{1.4}
    \setlength{\tabcolsep}{5.1pt}
    \begin{threeparttable}
        \scalebox{.65}{
            \begin{tabular}
                {   p{2cm}<{\centering}|
                p{2.2cm}<{\centering}|
                p{2.2cm}<{\centering}|
                p{2.2cm}<{\centering}|
                p{2.2cm}<{\centering}|
                p{2.2cm}<{\centering}|
                p{2.2cm}<{\centering}|
                p{2.2cm}<{\centering}
                }\hline\hline
                Dataset    & \multicolumn{7}{|c}{Diginetica}                                                                                                                                       \\ \hline
                Metric     & precision$\uparrow$                     & NDCG$\uparrow$    & MRR$\uparrow$     & $F_1$ $\uparrow$    & $wv$$\downarrow$  & rewarde$\uparrow$  & time$\downarrow$     \\ \hline
                w/ SCR     & {3.26$_{{(.014)}}$}                     & 13.5$_{{(.017)}}$ & 12.6$_{{(.023)}}$ & 5.43$_{{(.021)}}$   & 2.00$_{{(.013)}}$ & -2.08$_{{(.031)}}$ & 69126$_{{(.023)}}$   \\ \hline
                w/ IFRQE++ & {3.14}$_{{(.018)}}$                     & 12.5$_{{(.010)}}$ & 11.5$_{{(.012)}}$ & {5.23}$_{{(.028)}}$ & 2.01$_{{(.036)}}$ & -2.08$_{{(.032)}}$ & 739$_{{(.026)}}$     \\\hline \hline
                w/ SCR     & {2.09$_{{(.013)}}$}                     & 7.38$_{{(.012)}}$ & 6.37$_{{(.013)}}$ & 3.48$_{{(.011)}}$   & 2.00$_{{(.023)}}$ & -2.10$_{{(.011)}}$ & 21342$_{{(.033)}}$   \\ \hline
                w/ IFRQE++ & {1.82}$_{{(.028)}}$                     & 6.04$_{{(.018)}}$ & 5.04$_{{(.022)}}$ & {3.03}$_{{(.009)}}$ & 2.00$_{{(.016)}}$ & -2.09$_{{(.032)}}$ & 1443$_{{(.026)}}$    \\\hline \hline
                w/ SCR     & {1.19$_{{(.004)}}$}                     & 3.77$_{{(.027)}}$ & 3.05$_{{(.013)}}$ & 1.98$_{{(.012)}}$   & 1.99$_{{(.023)}}$ & -2.21$_{{(.011)}}$ & 21555$_{{(.023)}}$   \\ \hline
                w/ IFRQE++ & {0.84}$_{{(.014)}}$                     & 2.60$_{{(.012)}}$ & 2.07$_{{(.032)}}$ & {1.40}$_{{(.021)}}$ & 2.01$_{{(.032)}}$ & -2.37$_{{(.032)}}$ & 1331$_{{(.025)}}$    \\\hline \hline
                w/ SCR     & {1.56$_{{(.014)}}$}                     & 4.70$_{{(.029)}}$ & 3.68$_{{(.013)}}$ & 2.60$_{{(.019)}}$   & 1.97$_{{(.017)}}$ & -2.22$_{{(.037)}}$ & 137494$_{{(.033)}}$  \\ \hline
                w/ IFRQE++ & {1.11}$_{{(.016)}}$                     & 3.28$_{{(.015)}}$ & 2.54$_{{(.013)}}$ & {1.85}$_{{(.008)}}$ & 2.00$_{{(.026)}}$ & -2.23$_{{(.013)}}$ & 2328$_{{(.023)}}$    \\\hline \hline
                w/ SCR     & {0.86$_{{(.013)}}$}                     & 2.51$_{{(.019)}}$ & 1.94$_{{(.025)}}$ & 1.43$_{{(.021)}}$   & 2.00$_{{(.017)}}$ & -2.06$_{{(.010)}}$ & 7320$_{{(.032)}}$    \\ \hline
                w/ IFRQE++ & {0.85}$_{{(.018)}}$                     & 2.54$_{{(.020)}}$ & 1.99$_{{(.022)}}$ & {1.41}$_{{(.023)}}$ & 2.01$_{{(.016)}}$ & -2.07$_{{(.032)}}$ & 879$_{{(.033)}}$     \\\hline \hline
                Dataset    & \multicolumn{7}{|c}{Amazon Video Games}                                                                                                                               \\ \hline
                Metric     & precision$\uparrow$                     & NDCG$\uparrow$    & MRR$\uparrow$     & $F_1$ $\uparrow$    & $wv$$\downarrow$  & rewarde$\uparrow$  & time$\downarrow$     \\ \hline
                w/ SCR     & {1.25$_{{(.022)}}$}                     & 4.50$_{{(.027)}}$ & 3.92$_{{(.009)}}$ & 2.08$_{{(.011)}}$   & 2.32$_{{(.023)}}$ & -2.40$_{{(.031)}}$ & 9415$_{{(.023)}}$    \\ \hline
                w/ IFRQE++ & {1.09}$_{{(.018)}}$                     & 3.79$_{{(.010)}}$ & 3.24$_{{(.021)}}$ & {1.82}$_{{(.023)}}$ & 2.35$_{{(.016)}}$ & -2.42$_{{(.012)}}$ & 1021$_{{(.036)}}$    \\\hline \hline
                w/ SCR     & {1.09$_{{(.007)}}$}                     & 3.34$_{{(.017)}}$ & 2.65$_{{(.013)}}$ & 1.81$_{{(.021)}}$   & 2.32$_{{(.023)}}$ & -2.52$_{{(.011)}}$ & 14943$_{{(.033)}}$   \\ \hline
                w/ IFRQE++ & {0.86}$_{{(.028)}}$                     & 2.64$_{{(.011)}}$ & 2.09$_{{(.012)}}$ & {1.43}$_{{(.038)}}$ & 2.35$_{{(.016)}}$ & -2.50$_{{(.023)}}$ & 3192$_{{(.026)}}$    \\\hline \hline
                w/ SCR     & {1.23$_{{(.0245}}$}                     & 4.43$_{{(.027)}}$ & 3.87$_{{(.015)}}$ & 2.04$_{{(.022)}}$   & 2.31$_{{(.023)}}$ & -2.68$_{{(.031)}}$ & 18492$_{{(.023)}}$   \\ \hline
                w/ IFRQE++ & {1.19}$_{{(.019)}}$                     & 4.29$_{{(.014)}}$ & 3.74$_{{(.019)}}$ & {1.98}$_{{(.018)}}$ & 2.34$_{{(.031)}}$ & -2.63$_{{(.019)}}$ & 2148$_{{(.026)}}$    \\\hline \hline
                w/ SCR     & {1.57$_{{(.029)}}$}                     & 5.71$_{{(.017)}}$ & 5.31$_{{(.015)}}$ & 3.07$_{{(.022)}}$   & 2.00$_{{(.023)}}$ & -2.07$_{{(.031)}}$ & 161501$_{{(.033)}}$  \\ \hline
                w/ IFRQE++ & {1.59}$_{{(.019)}}$                     & 5.94$_{{(.011)}}$ & 5.59$_{{(.022)}}$ & {3.11}$_{{(.021)}}$ & 2.01$_{{(.016)}}$ & -2.06$_{{(.023)}}$ & 2743$_{{(.013)}}$    \\\hline \hline
                w/ SCR     & {0.97$_{{(.025)}}$}                     & 2.97$_{{(.019)}}$ & 2.35$_{{(.029)}}$ & 1.61$_{{(.009)}}$   & 2.31$_{{(.013)}}$ & -2.52$_{{(.021)}}$ & 1045387$_{{(.023)}}$ \\ \hline
                w/ IFRQE++ & {1.13}$_{{(.010)}}$                     & 3.64$_{{(.017)}}$ & 2.99$_{{(.011)}}$ & {1.88}$_{{(.021)}}$ & 2.34$_{{(.016)}}$ & -2.44$_{{(.027)}}$ & 579$_{{(.013)}}$     \\\hline \hline
            \end{tabular}
        }
    \end{threeparttable}
    \label{tab:ex2}
\end{table*}

\subsection{More Experiments}
In this section, we present more experiments to evaluate and analyze our proposed frameworks.

\subsubsection{Comparison between our framework and the method of ``training from scratch''}
In this section, we compare our framework with the method of ``training from scratch'' (we call it as \textbf{SCR}), where we drop the influence function, and for each action exploration, we retrain the recommender model. We remain the other model components of this method the same as our framework.
The comparison results are presented in Table~\ref{tab:ex2}.
We can see, the performances of our framework do not surpass SCR in most cases.
This is understandable, since SCR uses the true loss, and our framework only leverages the approximated values.
However, we find that the reward gap is not large, which may suggest that our designed influence function can well approximate the true loss, and help to achieve satisfied reward.
An important superiority of our framework is the efficiency. As can be seen in the last column of Table~\ref{tab:ex2}, we can improve the training efficiency by about 20.3 times. This superiority is very important for the recommender system, which is an on-line service, and has to make quick responses the user feedback.

\begin{table*}[h]
    \caption{\small{Comparison between different models with different $\lambda$'s. We use bold fonts to label the best performance for each dataset, evaluation metric and base model. ``()'' indicates the standard error.
        }
    }
    \center
    \renewcommand\arraystretch{1.3}
    \setlength{\tabcolsep}{5.pt}
    \begin{threeparttable}
        \scalebox{.5}{
            \begin{tabular}
                { p{6cm}<{\centering}|
                p{4.5cm}<{\centering}|
                p{4.5cm}<{\centering}|
                p{4.5cm}<{\centering}|
                p{4.5cm}<{\centering}
                } \hline\hline
                Dataset                         &
                \multicolumn{1}{c|}{Simulation} &
                \multicolumn{1}{c|}{Diginetica} &
                \multicolumn{1}{c|}{Steam}      &
                \multicolumn{1}{c}{Amazon Video}                              \\ \hline
                \multicolumn{5}{c}{$\lambda=0.1$}                             \\\hline
                \makecell[c]{MF}                & -0.34$_{{(.002)}}$
                                                & -0.29$_{{(.009)}}$
                                                & -0.49$_{{(.005)}}$
                                                & -0.32$_{{(.007)}}$          \\ \hline
                w/ Random
                                                & -0.46$_{{(.005)}}$
                                                & -0.28$_{{(.017)}}$
                                                & -0.48$_{{(.019)}}$
                                                & -0.31$_{{(.011)}}$          \\
                w/ Threshold
                                                & -0.48$_{{(.015)}}$
                                                & -0.48$_{{(.022)}}$
                                                & -0.50$_{{(.017)}}$
                                                & -0.30$_{{(.018)}}$          \\
                w/ Proactive
                                                & -0.33$_{{(.034)}}$
                                                & -0.28$_{{(.025)}}$
                                                & -0.31$_{{(.007)}}$
                                                & -0.36$_{{(.019)}}$          \\
                w/ IFRQE
                                                & -0.34$_{{(.007)}}$
                                                & -0.23$_{{(.011)}}$
                                                & -0.49$_{{(.013)}}$
                                                & -0.30$_{{(.008)}}$          \\

                w/ IFRQE++                      & \textbf{-0.32}$_{{(.006)}}$
                                                & \textbf{-0.22}$_{{(.005)}}$
                                                & \textbf{-0.47}$_{{(.006)}}$
                                                & \textbf{-0.28}$_{{(.002)}}$ \\\hline \hline

                \makecell[c]{NeuMF}             & -0.40$_{{(.001)}}$
                                                & -0.30$_{{(.002)}}$
                                                & -0.37$_{{(.009)}}$
                                                & -0.33$_{{(.007)}}$          \\ \hline

                w/ Random                       & -0.39$_{{(.012)}}$
                                                & -0.31$_{{(.010)}}$
                                                & -0.36$_{{(.015)}}$
                                                & -0.34$_{{(.006)}}$          \\
                w/ Threshold
                                                & -0.38$_{{(.019)}}$
                                                & -0.38$_{{(.012)}}$
                                                & -0.35$_{{(.007)}}$
                                                & -0.41$_{{(.014)}}$          \\
                w/ Proactive
                                                & -0.49$_{{(.027)}}$
                                                & -0.35$_{{(.012)}}$
                                                & -0.37$_{{(.007)}}$
                                                & -0.35$_{{(.019)}}$          \\
                w/ IFRQE                        & -0.37$_{{(.005)}}$
                                                & -0.34$_{{(.002)}}$
                                                & -0.25$_{{(.014)}}$
                                                & -0.37$_{{(.012)}}$          \\

                w/ IFRQE++                      & \textbf{-0.34}$_{{(.007)}}$
                                                & \textbf{-0.28}$_{{(.003)}}$
                                                & \textbf{-0.24}$_{{(.005)}}$
                                                & \textbf{-0.32}$_{{(.002)}}$ \\\hline \hline

                \makecell[c]{LightGCN}          & -0.32$_{{(.011)}}$
                                                & -0.28$_{{(.016)}}$
                                                & -0.36$_{{(.013)}}$
                                                & -0.49$_{{(.011)}}$          \\ \hline

                w/ Random                       & -0.30$_{{(.013)}}$
                                                & -0.27$_{{(.022)}}$
                                                & -0.39$_{{(.019)}}$
                                                & -0.47$_{{(.016)}}$          \\
                w/ Threshold
                                                & -0.36$_{{(.021)}}$
                                                & -0.33$_{{(.030)}}$
                                                & -0.35$_{{(.026)}}$
                                                & -0.34$_{{(.011)}}$          \\
                w/ Proactive
                                                & -0.29$_{{(.034)}}$
                                                & -0.51$_{{(.025)}}$
                                                & -0.34$_{{(.007)}}$
                                                & -0.25$_{{(.019)}}$          \\
                w/ IFRQE                        & -0.26$_{{(.005)}}$
                                                & -0.27$_{{(.014)}}$
                                                & -0.34$_{{(.011)}}$
                                                & -0.25$_{{(.003)}}$          \\

                w/ IFRQE++                      & \textbf{-0.25}$_{{(.006)}}$
                                                & \textbf{-0.25}$_{{(.016)}}$
                                                & \textbf{-0.33}$_{{(.008)}}$
                                                & \textbf{-0.24}$_{{(.004)}}$ \\\hline\hline

                \multicolumn{5}{c}{$\lambda=0.5$}                             \\\hline
                \makecell[c]{MF}                & -1.19$_{{(.006)}}$
                                                & -1.10$_{{(.012)}}$
                                                & -1.54$_{{(.011)}}$
                                                & -1.27$_{{(.016)}}$          \\ \hline
                w/ Random
                                                & -1.28$_{{(.016)}}$
                                                & -1.08$_{{(.007)}}$
                                                & -1.52$_{{(.019)}}$
                                                & -1.21$_{{(.018)}}$          \\
                w/ Threshold
                                                & -1.32$_{{(.021)}}$
                                                & -1.44$_{{(.030)}}$
                                                & -1.37$_{{(.026)}}$
                                                & -1.21$_{{(.011)}}$          \\
                w/ Proactive
                                                & -1.18$_{{(.028)}}$
                                                & -1.08$_{{(.025)}}$
                                                & -1.44$_{{(.007)}}$
                                                & -0.36$_{{(.019)}}$          \\
                w/ IFRQE
                                                & -0.97$_{{(.003)}}$
                                                & -0.85$_{{(.007)}}$
                                                & -0.77$_{{(.006)}}$
                                                & -1.04$_{{(.008)}}$          \\

                w/ IFRQE++                      & \textbf{-0.95}$_{{(.001)}}$
                                                & \textbf{-0.81}$_{{(.006)}}$
                                                & \textbf{-0.76}$_{{(.001)}}$
                                                & \textbf{-1.03}$_{{(.004)}}$ \\\hline \hline

                \makecell[c]{NeuMF}             & -1.11$_{{(.009)}}$
                                                & -1.11$_{{(.002)}}$
                                                & -1.42$_{{(.006)}}$
                                                & -1.27$_{{(.009)}}$          \\ \hline

                w/ Random                       & -1.22$_{{(.006)}}$
                                                & -1.10$_{{(.010)}}$
                                                & -1.40$_{{(.013)}}$
                                                & -1.22$_{{(.017)}}$          \\
                w/ Threshold
                                                & -1.32$_{{(.021)}}$
                                                & -1.44$_{{(.030)}}$
                                                & -1.37$_{{(.026)}}$
                                                & -1.23$_{{(.011)}}$          \\
                w/ Proactive
                                                & -1.21$_{{(.017)}}$
                                                & -1.29$_{{(.025)}}$
                                                & -1.30$_{{(.007)}}$
                                                & -0.36$_{{(.019)}}$          \\
                w/ IFRQE                        & -0.93$_{{(.011)}}$
                                                & -1.01$_{{(.002)}}$
                                                & -1.30$_{{(.005)}}$
                                                & -1.26$_{{(.004)}}$          \\

                w/ IFRQE++                      & \textbf{-0.91}$_{{(.006)}}$
                                                & \textbf{-0.98}$_{{(.003)}}$
                                                & \textbf{-1.28}$_{{(.011)}}$
                                                & \textbf{-1.20}$_{{(.007)}}$ \\\hline \hline

                \makecell[c]{LightGCN}          & -1.17$_{{(.011)}}$
                                                & -1.09$_{{(.016)}}$
                                                & -1.41$_{{(.009)}}$
                                                & -1.48$_{{(.006)}}$          \\ \hline

                w/ Random                       & -1.11$_{{(.016)}}$
                                                & -1.06$_{{(.022)}}$
                                                & -1.43$_{{(.012)}}$
                                                & -1.38$_{{(.019)}}$          \\
                w/ Threshold
                                                & -1.32$_{{(.021)}}$
                                                & -1.44$_{{(.030)}}$
                                                & -1.74$_{{(.026)}}$
                                                & -1.75$_{{(.011)}}$          \\
                w/ Proactive
                                                & -1.20$_{{(.034)}}$
                                                & -1.54$_{{(.033)}}$
                                                & -1.19$_{{(.007)}}$
                                                & -1.70$_{{(.019)}}$          \\
                w/ IFRQE
                                                & -1.04$_{{(.007)}}$
                                                & -0.97$_{{(.014)}}$
                                                & -1.41$_{{(.003)}}$
                                                & -1.33$_{{(.006)}}$          \\

                w/ IFRQE++                      & \textbf{-1.02}$_{{(.001)}}$
                                                & \textbf{-0.96}$_{{(.016)}}$
                                                & \textbf{-1.40}$_{{(.008)}}$
                                                & \textbf{-1.29}$_{{(.005)}}$ \\\hline\hline

                \multicolumn{5}{c}{$\lambda=1.0$}                             \\\hline
                \makecell[c]{MF}                & -2.25$_{{(.032)}}$
                                                & -2.65$_{{(.072)}}$
                                                & -2.99$_{{(.093)}}$
                                                & -2.43$_{{(.022)}}$          \\ \hline
                w/ Random
                                                & -2.13$_{{(.014)}}$
                                                & -2.04$_{{(.006)}}$
                                                & -2.82$_{{(.015)}}$
                                                & -2.30$_{{(.023)}}$          \\
                w/ Threshold
                                                & -2.20$_{{(.011)}}$
                                                & -2.06$_{{(.013)}}$
                                                & -2.57$_{{(.023)}}$
                                                & -2.20$_{{(.015)}}$          \\
                w/ Proactive
                                                & -2.25$_{{(.034)}}$
                                                & -2.14$_{{(.012)}}$
                                                & -2.43$_{{(.007)}}$
                                                & -2.08$_{{(.019)}}$          \\
                w/ IFRQE
                                                & -2.09$_{{(.017)}}$
                                                & -2.18$_{{(.006)}}$
                                                & -2.43$_{{(.015)}}$
                                                & -2.10$_{{(.005)}}$          \\

                w/ IFRQE++                      & \textbf{-1.92}$_{{(.012)}}$
                                                & \textbf{-1.92}$_{{(.022)}}$
                                                & \textbf{-2.42}$_{{(.014)}}$
                                                & \textbf{-1.98}$_{{(.056)}}$ \\\hline \hline

                \makecell[c]{NeuMF}             & -2.32$_{{(.011)}}$
                                                & -2.11$_{{(.003)}}$
                                                & -2.76$_{{(.007)}}$
                                                & -2.47$_{{(.010)}}$          \\ \hline

                w/ Random                       & -2.19$_{{(.011)}}$
                                                & -2.11$_{{(.016)}}$
                                                & -2.59$_{{(.011)}}$
                                                & -2.41$_{{(.010)}}$          \\
                w/ Threshold
                                                & -2.18$_{{(.012)}}$
                                                & -2.08$_{{(.013)}}$
                                                & -2.86$_{{(.011)}}$
                                                & -2.21$_{{(.025)}}$          \\
                w/ Proactive
                                                & -2.18$_{{(.034)}}$
                                                & -2.14$_{{(.015)}}$
                                                & -2.42$_{{(.022)}}$
                                                & -2.41$_{{(.019)}}$          \\
                w/ IFRQE                        & -2.48$_{{(.013)}}$
                                                & -2.17$_{{(.019)}}$
                                                & -2.45$_{{(.027)}}$
                                                & -2.24$_{{(.012)}}$          \\

                w/ IFRQE++                      & \textbf{-2.11}$_{{(.011)}}$
                                                & \textbf{-2.08}$_{{(.023)}}$
                                                & \textbf{-2.37}$_{{(.027)}}$
                                                & \textbf{-2.17}$_{{(.011)}}$ \\\hline \hline

                \makecell[c]{LightGCN}          & -2.82$_{{(.012)}}$
                                                & -2.71$_{{(.009)}}$
                                                & -3.13$_{{(.014)}}$
                                                & -3.05$_{{(.007)}}$          \\ \hline

                w/ Random                       & -2.71$_{{(.022)}}$
                                                & -2.64$_{{(.007)}}$
                                                & -3.07$_{{(.019)}}$
                                                & -2.93$_{{(.015)}}$          \\
                w/ Threshold
                                                & -2.70$_{{(.008)}}$
                                                & -2.60$_{{(.031)}}$
                                                & -2.82$_{{(.022)}}$
                                                & -2.30$_{{(.021)}}$          \\
                w/ Proactive
                                                & -2.47$_{{(.034)}}$
                                                & -2.11$_{{(.013)}}$
                                                & -2.81$_{{(.027)}}$
                                                & -2.64$_{{(.039)}}$          \\
                w/ IFRQE
                                                & -2.82$_{{(.014)}}$
                                                & -2.65$_{{(.013)}}$
                                                & -2.82$_{{(.008)}}$
                                                & -2.13$_{{(.013)}}$          \\

                w/ IFRQE++                      & \textbf{-2.44}$_{{(.016)}}$
                                                & \textbf{-2.55}$_{{(.009)}}$
                                                & \textbf{-2.80}$_{{(.005)}}$
                                                & \textbf{-2.03}$_{{(.023)}}$ \\\hline\hline

                \multicolumn{5}{c}{$\lambda=2.0$}                             \\\hline
                \makecell[c]{MF}                & -4.53$_{{(.011)}}$
                                                & -4.23$_{{(.052)}}$
                                                & -5.46$_{{(.013)}}$
                                                & -4.56$_{{(.012)}}$          \\ \hline
                w/ Random
                                                & -4.30$_{{(.024)}}$
                                                & -4.01$_{{(.016)}}$
                                                & -5.18$_{{(.025)}}$
                                                & -4.64$_{{(.023)}}$          \\
                w/ Threshold
                                                & -4.34$_{{(.009)}}$
                                                & -3.72$_{{(.027)}}$
                                                & -4.92$_{{(.023)}}$
                                                & -4.32$_{{(.025)}}$          \\
                w/ Proactive
                                                & -4.37$_{{(.014)}}$
                                                & -4.10$_{{(.021)}}$
                                                & -4.55$_{{(.027)}}$
                                                & -4.09$_{{(.029)}}$          \\
                w/ IFRQE
                                                & -4.12$_{{(.017)}}$
                                                & -3.24$_{{(.026)}}$
                                                & -4.05$_{{(.015)}}$
                                                & -3.71$_{{(.043)}}$          \\

                w/ IFRQE++                      & \textbf{-4.06}$_{{(.010)}}$
                                                & \textbf{-3.16}$_{{(.032)}}$
                                                & \textbf{-4.02}$_{{(.014)}}$
                                                & \textbf{-3.37}$_{{(.056)}}$ \\\hline \hline

                \makecell[c]{NeuMF}             & -4.44$_{{(.012)}}$
                                                & -4.12$_{{(.043)}}$
                                                & -5.34$_{{(.027)}}$
                                                & -4.81$_{{(.019)}}$          \\ \hline

                w/ Random                       & -4.21$_{{(.012)}}$
                                                & -4.05$_{{(.035)}}$
                                                & -5.07$_{{(.021)}}$
                                                & -4.58$_{{(.031)}}$          \\
                w/ Threshold
                                                & -4.38$_{{(.012)}}$
                                                & -3.74$_{{(.013)}}$
                                                & -4.81$_{{(.011)}}$
                                                & -4.33$_{{(.035)}}$          \\
                w/ Proactive
                                                & -4.65$_{{(.024)}}$
                                                & -3.77$_{{(.025)}}$
                                                & -4.62$_{{(.007)}}$
                                                & -4.18$_{{(.009)}}$          \\
                w/ IFRQE
                                                & -4.56$_{{(.013)}}$
                                                & -4.00$_{{(.019)}}$
                                                & -3.90$_{{(.028)}}$
                                                & -3.06$_{{(.022)}}$          \\

                w/ IFRQE++                      & \textbf{-4.00}$_{{(.012)}}$
                                                & \textbf{-3.67}$_{{(.021)}}$
                                                & \textbf{-3.40}$_{{(.019)}}$
                                                & \textbf{-2.74}$_{{(.012)}}$ \\\hline \hline

                \makecell[c]{LightGCN}          & -4.94$_{{(.021)}}$
                                                & -4.10$_{{(.009)}}$
                                                & -5.75$_{{(.024)}}$
                                                & -5.20$_{{(.057)}}$          \\ \hline

                w/ Random                       & -4.70$_{{(.032)}}$
                                                & -3.90$_{{(.007)}}$
                                                & -5.07$_{{(.029)}}$
                                                & -4.73$_{{(.033)}}$          \\
                w/ Threshold
                                                & -4.36$_{{(.017)}}$
                                                & -4.32$_{{(.031)}}$
                                                & -5.21$_{{(.022)}}$
                                                & -4.93$_{{(.041)}}$          \\
                w/ Proactive
                                                & -3.63$_{{(.035)}}$
                                                & -4.72$_{{(.039)}}$
                                                & -4.54$_{{(.007)}}$
                                                & -4.71$_{{(.011)}}$          \\
                w/ IFRQE
                                                & -3.89$_{{(.013)}}$
                                                & -2.54$_{{(.013)}}$
                                                & -3.90$_{{(.018)}}$
                                                & -3.71$_{{(.023)}}$          \\

                w/ IFRQE++                      & \textbf{-3.55}$_{{(.026)}}$
                                                & \textbf{-2.52}$_{{(.009)}}$
                                                & \textbf{-3.40}$_{{(.025)}}$
                                                & \textbf{-3.69}$_{{(.033)}}$ \\\hline\hline
            \end{tabular}
        }
    \end{threeparttable}
    \label{lb}
\end{table*}

\subsubsection{Influence of the Data Sparsity}
In real-world scenarios, recommender systems can be applied in different applications, where the dataset sparsity may vary a lot.
In this section, we would like to study whether our methods are consistently competitive for the datasets with different sparsities.
In order to flexibly control the sparsity, we conduct this experiment based on the simulation dataset.
Since the threshold $\eta$ controls the hardness of generating the user-item interactions, we tune $\eta$ in the range of $\{0.1,0.2,0.3,0.4,0.5\}$ to build the datasets with different densities, where larger $\eta$ can lead more sparse dataset.
The statistics of the generated datasets are presented in Table~\ref{sparse}.
In Figure~\ref{sp}, we report the performance of different models based on the reward, where we can see:
the performance of the base model is not satisfied in most cases.
IFRQE usually outperforms the random method, although there are a few exceptions.
IFRQE++ can always achieves the best performance, which is consistent on all the base models and datasets with different sparsities.
These results demonstrate the robustness of our model, and suggests that it can be potentially applied to a wide range of real-world applications.

\subsubsection{Influence of the balancing parameter $\lambda$}
In the reward function, $\lambda$ balances the importances of the recommendation quality and user disclosing willingness.
To study whether our model can adaptively trade-off the above two aspects, we specify $\lambda$ with different values, and observe whether our model can always achieve better performance than the baselines.
In specific, {we set $\lambda$ as 0.1, 0.5, 1.0 and 2.0 respectively, and the results of comparing our models with the baselines are presented in Table~\ref{lb}.}
We can see: on different datasets, because the base model completely ignores the user disclosing willingness, the overall reward is the worst comparing with the other methods.
Blindly integrating the user disclosing willingness is also sub-optimal, which is evidenced by the lower performance of the random method.
By designing a principled model to optimize the overall reward, IFRQE can achieve better performance than the base and random models in most cases.
As expected, by leveraging more anchor selection vectors to simulate the validation loss, the final model IFRQE++ achieves the best performance.
The above observations are consistent for different $\lambda$'s, which manifests that our model is robust to the predefined relative importance between the recommendation quality and user disclosing willingness.

\begin{figure}[t]
    \centering
    \setlength{\fboxrule}{0.pt}
    \setlength{\fboxsep}{0.pt}
    \fbox{
        \includegraphics[width=1.\linewidth]{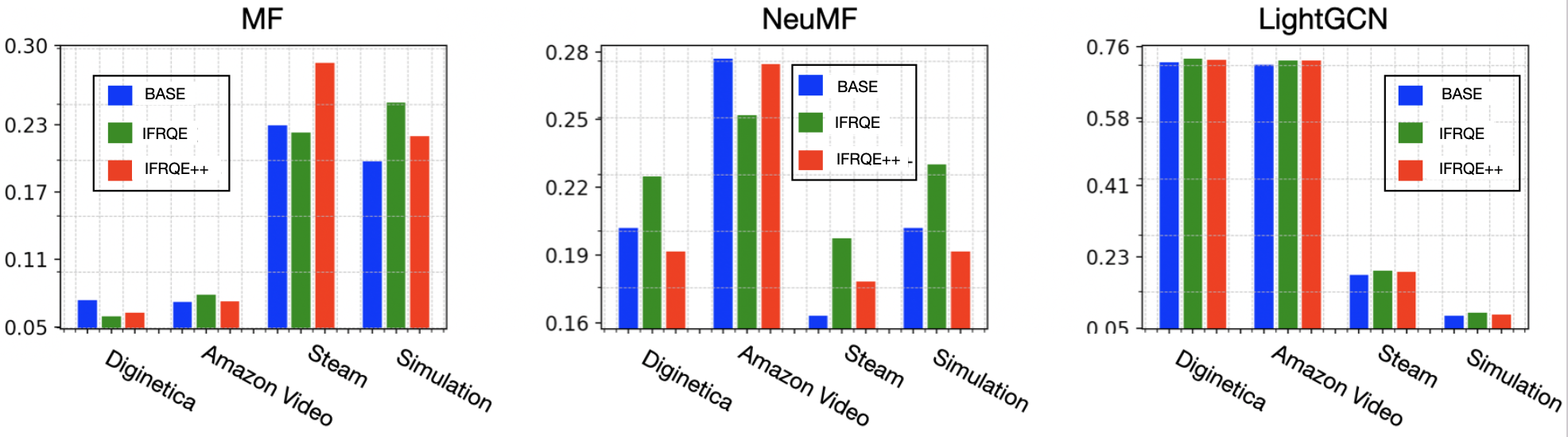}
    }
    \caption{Approximation error on the validation loss for all the datasets and base models. }
    \label{app1}
\end{figure}

\begin{figure}[t]
    \centering
    \setlength{\fboxrule}{0.pt}
    \setlength{\fboxsep}{0.pt}
    \fbox{
        \includegraphics[width=1.\linewidth]{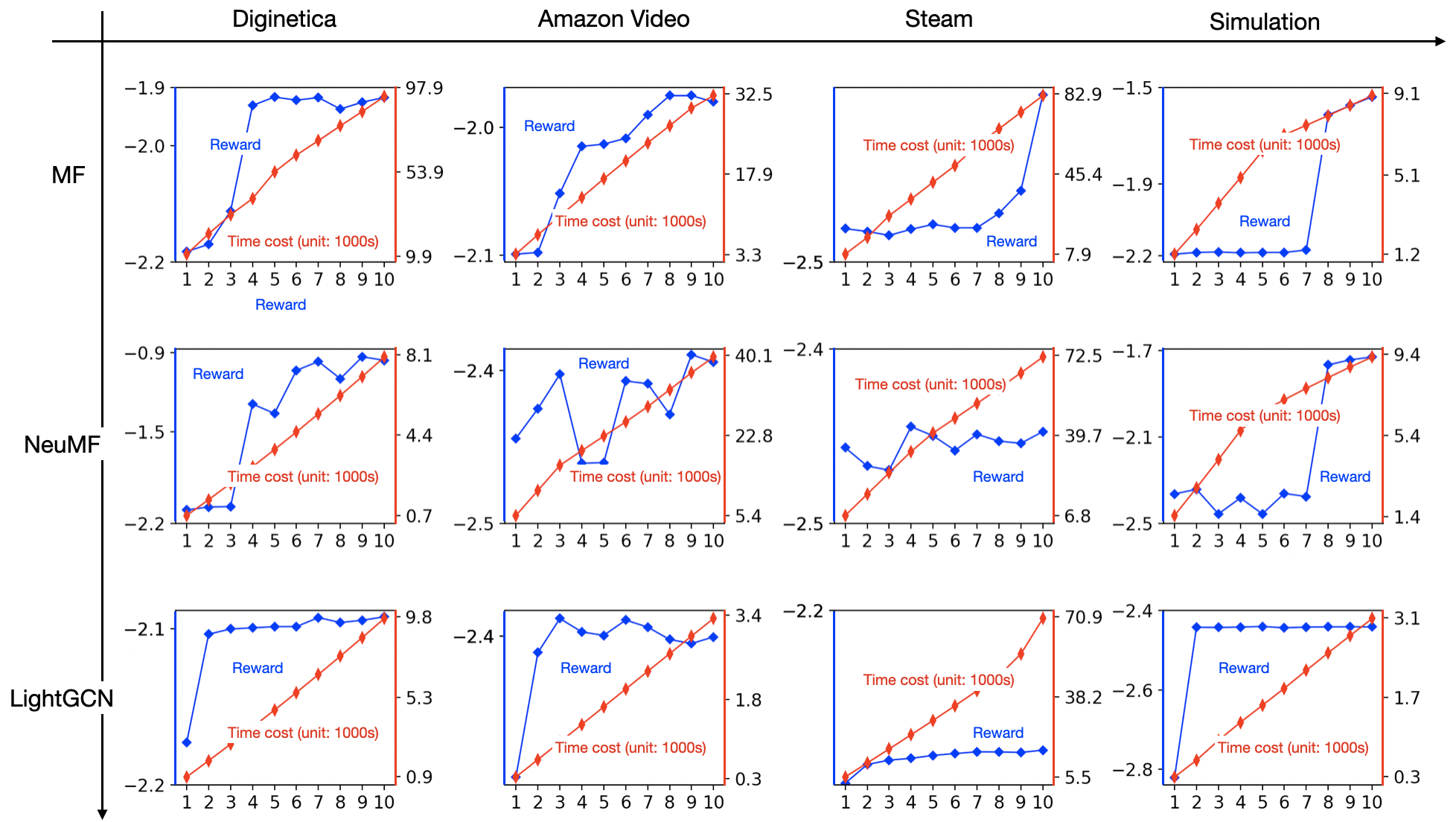}
    }
    \caption{Influence of $T$ for all the datasets and base models. }
    \label{app2}
\end{figure}

\begin{table*}[t]
    \caption{Additional metrics for evaluating the recommendation performance. We use ``P'' to represent the precision. All the results are percentage values with ``\%'' omitted.}
    \center
    \renewcommand\arraystretch{1.4}
    \setlength{\tabcolsep}{5.1pt}
    \begin{threeparttable}
        \scalebox{.65}{
            \begin{tabular}
                { l|
                p{1.1cm}<{\centering}p{1.1cm}<{\centering}p{1.5cm}<{\centering}|
                p{1.1cm}<{\centering}p{1.1cm}<{\centering}p{1.5cm}<{\centering}|
                p{1.1cm}<{\centering}p{1.1cm}<{\centering}p{1.5cm}<{\centering}|
                p{1.1cm}<{\centering}p{1.1cm}<{\centering}p{1.5cm}<{\centering}
                } \hline\hline
                Dataset                         &
                \multicolumn{3}{c|}{Simulation} &
                \multicolumn{3}{c|}{Diginetica} &
                \multicolumn{3}{c|}{Steam}      &
                \multicolumn{3}{c}{Amazon Video}                                                                                                                                                                                                                                     \\ \hline
                Metric                          & P$\uparrow$                & NDCG$\uparrow$             & MRR$\uparrow$               & P$\uparrow$ & NDCG$\uparrow$ & MRR$\uparrow$ & P$\uparrow$ & NDCG$\uparrow$ & MRR$\uparrow$ & P$\uparrow$ & NDCG$\uparrow$ & MRR$\uparrow$ \\ \hline
                \makecell[c]{MF}                & {0.39$_{{(.022)}}$}        & 1.02$_{{(.017)}}$          & 0.72$_{{(.023)}}$
                                                & 2.83$_{{(.012)}}$          & 11.3$_{{(.017)}}$          & 10.3$_{{(.023)}}$
                                                & 6.21$_{{(.043)}}$          & 19.5$_{{(.014)}}$          & 15.8$_{{(.093)}}$
                                                & \textbf{1.03}$_{{(.012)}}$ & \textbf{3.52}$_{{(.021)}}$ & \textbf{2.99}$_{{(.022)}}$                                                                                                                                               \\ \hline
                w/ Random
                                                & \textbf{0.41}$_{{(.031)}}$ & \textbf{1.11}$_{{(.056)}}$ & 0.80$_{{(.024)}}$
                                                & 2.51$_{{(.031)}}$          & 9.88$_{{(.006)}}$          & 9.00$_{{(.014)}}$
                                                & 5.64$_{{(.019)}}$          & 17.7$_{{(.017)}}$          & 14.3$_{{(.015)}}$
                                                & 0.92$_{{(.011)}}$          & 3.10$_{{(.022)}}$          & 2.61$_{{(.023)}}$                                                                                                                                                        \\
                w/ Threshold
                                                & 0.34$_{{(.036)}}$          & 1.07$_{{(.022)}}$          & \textbf{1.01}$_{{(.011)}}$
                                                & 2.80$_{{(.022)}}$          & 11.0$_{{(.013)}}$          & 9.98$_{{(.042)}}$
                                                & 5.55$_{{(.028)}}$          & 17.7$_{{(.033)}}$          & 14.4$_{{(.027)}}$
                                                & 0.90$_{{(.026)}}$          & 2.86$_{{(.037)}}$          & 2.33$_{{(.041)}}$                                                                                                                                                        \\
                w/ Proactive
                                                & 0.37$_{{(.015)}}$          & 1.01$_{{(.021)}}$          & 0.93$_{{(.013)}}$
                                                & 1.33$_{{(.005)}}$          & 4.73$_{{(.023)}}$          & 4.10$_{{(.042)}}$
                                                & 5.96$_{{(.011)}}$          & 19.4$_{{(.013)}}$          & 16.0$_{{(.009)}}$
                                                & 0.93$_{{(.015)}}$          & 3.15$_{{(.017)}}$          & 2.66$_{{(.031)}}$                                                                                                                                                        \\
                w/ IFRQE
                                                & 0.37$_{{(.025)}}$          & \textbf{1.11}$_{{(.033)}}$ & 0.87$_{{(.017)}}$
                                                & \textbf{3.14}$_{{(.025)}}$ & \textbf{12.5}$_{{(.033)}}$ & \textbf{11.5}$_{{(.017)}}$
                                                & \textbf{6.23$_{{(.012)}}$} & \textbf{19.8}$_{{(.012)}}$ & \textbf{16.1}$_{{(.015)}}$
                                                & 0.34$_{{(.015)}}$          & 1.28$_{{(.016)}}$          & 1.14$_{{(.005)}}$                                                                                                                                                        \\

                w/ IFRQE++                      & {0.33}$_{{(.027)}}$        & 1.02$_{{(.013)}}$          & 0.82$_{(.012)}$
                                                & {2.41}$_{{(.007)}}$        & 9.46$_{{(.013)}}$          & 8.21$_{{(.012)}}$
                                                & {5.90}$_{{(.017)}}$        & 18.9$_{{(.032)}}$          & 15.4$_{{(.014)}}$
                                                & \textbf{1.03}$_{{(.012)}}$ & {3.37}$_{{(.032)}}$        & 2.78$_{{(.056)}}$                                                                                                                                                        \\\hline \hline

                \makecell[c]{NeuMF}             & 0.52$_{{(.037)}}$          & 1.40$_{{(.003)}}$          & 1.02$_{{(.011)}}$
                                                & \textbf{2.53}$_{{(.037)}}$ & \textbf{10.5}$_{{(.003)}}$ & \textbf{9.83}$_{{(.011)}}$
                                                & \textbf{6.12}$_{{(.007)}}$ & \textbf{19.5}$_{{(.012)}}$ & \textbf{15.9}$_{{(.007)}}$
                                                & 0.99$_{{(.014)}}$          & 3.38$_{{(.013)}}$          & 2.87$_{{(.010)}}$                                                                                                                                                        \\ \hline

                w/ Random                       & 0.54$_{{(.019)}}$          & 1.71$_{{(.015)}}$          & 1.39$_{{(.012)}}$
                                                & 1.43$_{{(.019)}}$          & 5.04$_{{(.015)}}$          & 4.34$_{{(.012)}}$
                                                & 4.64$_{{(.005)}}$          & 14.6$_{{(.019)}}$          & 11.5$_{{(.011)}}$
                                                & {1.08}$_{{(.014)}}$        & 3.63$_{{(.006)}}$          & 3.04$_{{(.010)}}$                                                                                                                                                        \\
                w/ Threshold
                                                & 0.90$_{{(.022)}}$          & 4.02$_{{(.028)}}$          & 3.86$_{{(.011)}}$
                                                & 2.43$_{{(.011)}}$          & 9.74$_{{(.044)}}$          & 8.94$_{{(.016)}}$
                                                & 4.50$_{{(.029)}}$          & 14.2$_{{(.022)}}$          & 11.5$_{{(.021)}}$
                                                & 0.93$_{{(.012)}}$          & 3.03$_{{(.028)}}$          & 2.49$_{{(.017)}}$                                                                                                                                                        \\
                w/ Proactive
                                                & 1.10$_{{(.033)}}$          & 4.47$_{{(.022)}}$          & \textbf{4.13}$_{{(.012)}}$
                                                & 0.78$_{{(.017)}}$          & 2.31$_{{(.043)}}$          & 1.79$_{{(.011)}}$
                                                & 5.73$_{{(.028)}}$          & 17.8$_{{(.032)}}$          & 14.3$_{{(.012)}}$
                                                & 0.86$_{{(.006)}}$          & 2.82$_{{(.021)}}$          & 2.33$_{{(.039)}}$                                                                                                                                                        \\

                w/ IFRQE                        & 0.84$_{{(.014)}}$          & 2.73$_{{(.023)}}$          & 2.26$_{{(.033)}}$
                                                & 1.75$_{{(.004)}}$          & 7.27$_{{(.023)}}$          & 2.99$_{{(.033)}}$
                                                & 4.90$_{{(.023)}}$          & 14.6$_{{(.019)}}$          & 11.4$_{{(.027)}}$
                                                & \textbf{1.10}$_{{(.015)}}$ & \textbf{3.77}$_{{(.014)}}$ & \textbf{3.39}$_{{(.012)}}$                                                                                                                                               \\

                w/ IFRQE++                      & \textbf{1.80}$_{{(.015)}}$ & \textbf{6.34}$_{{(.003)}}$ & 3.77$_{{(.021)}}$
                                                & 1.80$_{{(.015)}}$          & 6.34$_{{(.003)}}$          & 3.77$_{{(.021)}}$
                                                & {5.32}$_{{(.021)}}$        & {16.7}$_{{(.014)}}$        & 13.4$_{{(.027)}}$
                                                & {0.72}$_{{(.031)}}$        & {2.65}$_{{(.014)}}$        & 2.48$_{{(.011)}}$                                                                                                                                                        \\\hline \hline

                \makecell[c]{LightGCN}          & 0.54$_{{(.025)}}$          & 1.41$_{{(.008)}}$          & 1.01$_{{(.042)}}$
                                                & \textbf{4.51}$_{{(.019)}}$ & \textbf{20.2}$_{{(.010)}}$ & \textbf{19.5} $_{{(.009)}}$
                                                & 6.04$_{{(.010)}}$          & 18.8$_{{(.009)}}$          & 15.1$_{{(.014)}}$
                                                & \textbf{1.29}$_{{(.021)}}$ & \textbf{4.90}$_{{(.008)}}$ & \textbf{4.40}$_{{(.007)}}$                                                                                                                                               \\ \hline

                w/ Random                       & 0.49$_{{(.013)}}$          & 1.27$_{{(.013)}}$          & 0.89$_{{(.022)}}$
                                                & 4.50$_{{(.013)}}$          & 20.0$_{{(.013)}}$          & 19.2$_{{(.022)}}$
                                                & 6.03$_{{(.027)}}$          & 18.8$_{{(.026)}}$          & 15.1$_{{(.019)}}$
                                                & 1.05$_{{(.003)}}$          & 3.84$_{{(.009)}}$          & 3.37$_{{(.01 5)}}$                                                                                                                                                       \\
                w/ Threshold
                                                & 0.58$_{{(.038)}}$          & 2.25$_{{(.018)}}$          & 2.04$_{{(.018)}}$
                                                & 2.43$_{{(.027)}}$          & 9.74$_{{(.041)}}$          & 8.94$_{{(.021)}}$
                                                & 5.79$_{{(.049)}}$          & 18.1$_{{(.027)}}$          & 14.6$_{{(.036)}}$
                                                & 0.97$_{{(.013)}}$          & 3.28$_{{(.023)}}$          & 2.78$_{{(.048)}}$                                                                                                                                                        \\
                w/ Proactive
                                                & \textbf{0.61}$_{{(.011)}}$ & \textbf{2.42}$_{{(.023)}}$ & \textbf{2.22}$_{{(.013)}}$
                                                & 3.31$_{{(.026)}}$          & 14.1$_{{(.014)}}$          & 13.3$_{{(.011)}}$
                                                & 6.21$_{{(.019)}}$          & \textbf{20.6}$_{{(.017)}}$ & \textbf{17.2}$_{{(.028)}}$
                                                & 0.77$_{{(.015)}}$          & 2.48$_{{(.021)}}$          & 2.04$_{{(.038)}}$                                                                                                                                                        \\

                w/ IFRQE                        & 0.58$_{{(.008)}}$          & 1.60$_{{(.010)}}$          & 1.18$_{{(.014)}}$
                                                & 2.93$_{{(.008)}}$          & 10.9$_{{(.010)}}$          & 9.74$_{{(.014)}}$
                                                & \textbf{6.21}$_{{(.016)}}$ & 19.8$_{{(.025)}}$          & \textbf{16.1}$_{{(.008)}}$
                                                & 1.14$_{{(.007)}}$          & 3.98$_{{(.018)}}$          & 3.42$_{{(.013)}}$                                                                                                                                                        \\

                w/ IFRQE++                      & {0.45}$_{{(.022)}}$        & 1.21$_{{(.008)}}$          & 0.87$_{{(.016)}}$
                                                & {3.16}$_{{(.022)}}$        & 12.1$_{{(.008)}}$          & 10.9$_{{(.016)}}$
                                                & {6.15}$_{{(.018)}}$        & 19.7$_{{(.018)}}$          & 16.0$_{{(.005)}}$
                                                & {0.97}$_{{(.013)}}$        & 3.44$_{{(.009)}}$          & 2.98$_{{(.023)}}$                                                                                                                                                        \\\hline\hline
                \makecell[c]{DIN}               & \textbf{0.86$_{{(.022)}}$} & \textbf{2.76}$_{{(.017)}}$ & \textbf{2.25}$_{{(.023)}}$
                                                & 2.48$_{{(.023)}}$          & 8.30$_{{(.005)}}$          & 6.95$_{{(.031)}}$
                                                & 7.01$_{{(.019)}}$          & 23.5$_{{(.048)}}$          & 19.7$_{{(.005)}}$
                                                & \textbf{1.38}$_{{(.038)}}$ & \textbf{5.38}$_{{(.012)}}$ & \textbf{4.39}$_{{(.026)}}$                                                                                                                                               \\ \hline

                w/ Random                       & 0.74$_{{(.031)}}$          & 2.30$_{{(.016)}}$          & 1.84$_{{(.024)}}$
                                                & \textbf{2.60}$_{{(.016)}}$ & \textbf{8.65}$_{{(.042)}}$ & \textbf{7.22}$_{{(.024)}}$
                                                & 6.54$_{{(.030)}}$          & 21.4$_{{(.042)}}$          & 17.7$_{{(.014)}}$
                                                & 1.27$_{{(.018)}}$          & 3.78$_{{(.023)}}$          & 2.95$_{{(.015)}}$                                                                                                                                                        \\
                w/ Threshold
                                                & 0.82$_{{(.004)}}$          & 2.30$_{{(.022)}}$          & 1.73$_{{(.029)}}$
                                                & 2.24$_{{(.010)}}$          & 7.45$_{{(.032)}}$          & 6.22$_{{(.036)}}$
                                                & 5.79$_{{(.036)}}$          & 24.9$_{{(.045)}}$          & 20.8$_{{(.029)}}$
                                                & 1.26$_{{(.009)}}$          & 3.84$_{{(.036)}}$          & 3.03$_{{(.022)}}$                                                                                                                                                        \\
                w/ Proactive
                                                & 0.73$_{{(.012)}}$          & 2.14$_{{(.024)}}$          & 1.65$_{{(.018)}}$
                                                & 1.64$_{{(.015)}}$          & 5.05$_{{(.017)}}$          & 4.03$_{{(.019)}}$
                                                & 7.60$_{{(.029)}}$          & 25.2$_{{(.035)}}$          & 21.1$_{{(.025)}}$
                                                & 0.98$_{{(.020)}}$          & 3.03$_{{(.021)}}$          & 2.43$_{{(.027)}}$                                                                                                                                                        \\
                w/ IFRQE                        & {0.72}$_{{(.027)}}$        & 2.48$_{{(.013)}}$          & 2.11$_{{(.012)}}$
                                                & 2.35$_{{(.017)}}$          & 8.04$_{{(.047)}}$          & 6.82$_{{(.022)}}$
                                                & \textbf{7.93}$_{{(.021)}}$ & \textbf{26.6}$_{{(.030)}}$ & \textbf{22.3}$_{{(.041)}}$
                                                & 0.74$_{{(.034)}}$          & 2.47$_{{(.019)}}$          & 2.04$_{{(.043)}}$                                                                                                                                                        \\

                w/ IFRQE++                      & {0.62}$_{{(.027)}}$        & 1.73$_{{(.013)}}$          & 1.27$_{{(.012)}}$
                                                & {2.10}$_{{(.014)}}$        & 6.78$_{{(.042)}}$          & 5.56$_{{(.030)}}$
                                                & {5.72}$_{{(.030)}}$        & 17.5$_{{(.016)}}$          & 13.9$_{{(.045)}}$
                                                & {1.09}$_{{(.021)}}$        & 3.47$_{{(.047)}}$          & 2.83$_{{(.038)}}$                                                                                                                                                        \\\hline\hline
                \makecell[c]{CDAE}              & \textbf{0.74}$_{{(.010)}}$ & 2.73$_{{(.015)}}$          & 2.41$_{{(.031)}}$
                                                & \textbf{0.89}$_{{(.009)}}$ & 2.65$_{{(.016)}}$          & 2.06$_{{(.042)}}$
                                                & 6.91$_{{(.008)}}$          & 21.8$_{{(.029)}}$          & 17.6$_{{(.014)}}$
                                                & 0.67$_{{(.011)}}$          & 2.01$_{{(.027)}}$          & 1.58$_{{(.045)}}$                                                                                                                                                        \\ \hline

                w/ Random                       & \textbf{0.74}$_{{(.013)}}$ & \textbf{2.87}$_{{(.013)}}$ & \textbf{2.60}$_{{(.022)}}$
                                                & 0.83$_{{(.014)}}$          & 2.63$_{{(.042)}}$          & \textbf{2.13}$_{{(.030)}}$
                                                & \textbf{6.96}$_{{(.021)}}$ & \textbf{22.0}$_{{(.003)}}$ & \textbf{17.8}$_{{(.041)}}$
                                                & 0.73$_{{(.011)}}$          & 2.27$_{{(.015)}}$          & 1.82$_{{(.031)}}$                                                                                                                                                        \\
                w/ Threshold
                                                & 0.36$_{{(.021)}}$          & 1.05$_{{(.003)}}$          & 0.81$_{{(.041)}}$
                                                & 0.83$_{{(.030)}}$          & 2.61$_{{(.026)}}$          & 2.11$_{{(.006)}}$
                                                & 5.34$_{{(.045)}}$          & 16.9$_{{(.044)}}$          & 13.8$_{{(.010)}}$
                                                & 0.73$_{{(.016)}}$          & 2.27$_{{(.040)}}$          & 1.82$_{{(.015)}}$                                                                                                                                                        \\
                w/ Proactive
                                                & 0.98$_{{(.014)}}$          & 3.03$_{{(.024)}}$          & 2.43$_{{(.028)}}$
                                                & 0.71$_{{(.017)}}$          & 2.11$_{{(.039)}}$          & 1.64$_{{(.007)}}$
                                                & 6.76$_{{(.026)}}$          & 21.2$_{{(.033)}}$          & 17.1$_{{(.014)}}$
                                                & 0.98$_{{(.012)}}$          & 3.03$_{{(.031)}}$          & 2.43$_{{(.022)}}$                                                                                                                                                        \\
                w/ IFRQE                        & 0.56$_{{(.011)}}$          & 0.87$_{{(.015)}}$          & 0.69$_{{(.027)}}$
                                                & \textbf{0.89}$_{{(.011)}}$ & \textbf{2.68}$_{{(.032)}}$ & 2.11$_{{(.013)}}$
                                                & 6.77$_{{(.008)}}$          & 21.2$_{{(.025)}}$          & 17.1$_{{(.023)}}$
                                                & 1.08$_{{(.027)}}$          & {3.62}$_{{(.029)}}$        & \textbf{3.03}$_{{(.046)}}$                                                                                                                                               \\

                w/ IFRQE++                      & {0.52}$_{{(.037)}}$        & 0.91$_{{(.005)}}$          & 0.79$_{{(.012)}}$
                                                & {0.86}$_{{(.007)}}$        & 2.58$_{{(.019)}}$          & 2.03$_{{(.042)}}$
                                                & {6.85}$_{{(.018)}}$        & 21.4$_{{(.046)}}$          & 17.2$_{{(.032)}}$
                                                & \textbf{1.10}$_{{(.027)}}$ & \textbf{3.63}$_{{(.029)}}$ & 3.00$_{{(.046)}}$                                                                                                                                                        \\\hline\hline
            \end{tabular}
        }
    \end{threeparttable}
    \label{tab:metric}
\end{table*}

\subsubsection{Complete results for section 5.3 and 5.4 in the main paper}
To begin with, we present the complete results of the experiments in section 5.3 of the main paper.
From the results shown in Figure~\ref{app1}, we can see: similar to the results in the main paper, the validation loss can be in general well approximated in most cases.
IFRQE++ can achieve better approximation accuracy than IFRQE, which demonstrate the effectiveness of using more anchor vectors for computing the validation loss.
In Figure~\ref{app2}, we show the complete results of the experiments in section 5.4.
We can see: the reward changing patterns seem to be quite diverse as more anchor selection vectors are leveraged in our model.
For example, in the case of MF + Diginetica, the reward has a performance jump from $T=3$ to $T=4$.
Similar performance jumping patterns can also be observed in the settings of NeuMF + Diginetica, LightGCN + Diginetica, LightGCN + Amazon Video and the simulation dataset.
However, in the case of NeuMF + Amazon Video, the reward changes irregularly as $T$ becomes larger.
While different combinations between the base model and dataset may lead to various performance change patterns, a common phenomenon is that, in most cases, the performance tends to be better as more anchor selection vectors.
Simultaneously, the time cost is increased almost linearly as more anchor vectors are deployed to achieve better performance.
These observations are aligned with the conclusions in the main paper.

\subsubsection{Overall comparison with more performance evaluation}
{To begin with, we augment Table 1 in the main paper by reporting the recommendation performance based on Precision, NDCG and MRR.
    From the results shown in Table~\ref{tab:metric}, we can draw similar conclusions as Table 1, that is, there are many cases that, although we have removed some items due to the user willingness, the recommendation performances are not lowered.
}

\end{document}